\documentclass[aps,prd,twocolumn,superscriptaddress]{revtex4}
\usepackage{epsfig,epsf}
\usepackage{amsmath}
\usepackage{amsthm}
\usepackage{amsfonts}
\usepackage{amssymb}
\usepackage{dsfont}
\usepackage{multirow}
\usepackage{appendix}
\usepackage{slashed}
\usepackage[active]{srcltx}
\usepackage{psfrag}

\begin{document}

\title{Interpretation of the new $\Omega_c^{0}$ states via their mass and
width}
\date{\today}
\author{S.~S.~Agaev}
\affiliation{Institute for Physical Problems, Baku State University, Az--1148 Baku,
Azerbaijan}
\author{K.~Azizi}
\affiliation{Department of Physics, Do\v{g}u\c{s} University, Acibadem-Kadik\"{o}y, 34722
Istanbul, Turkey}
\author{H.~Sundu}
\affiliation{Department of Physics, Kocaeli University, 41380 Izmit, Turkey}

\begin{abstract}
The mass and pole residue of the ground and first radially excited $%
\Omega_c^{0}$ states with spin-parities $J^{P}=1/2^{+},\,3/2^{+}$, as well
as P-wave $\Omega_c^{0}$ with $J^{P}=1/2^{-},\,3/2^{-}$ are calculated by
means of the two-point QCD sum rules. The strong decays of $\Omega_c^{0}$
baryons are also studied and width of these decay channels are computed. The
relevant computations are performed in the context of the full QCD sum rules
on the light-cone. Obtained results for the masses and widths are confronted
with recent experimental data of LHCb Collaboration, which allow us to
interpret $\Omega_c(3000)^{0}$, $\Omega_c(3050)^{0}$, and $%
\Omega_c(3119)^{0} $ as the excited $css$ baryons with the quantum numbers $%
(1P,\,1/2^{-})$, $(1P,\,3/2^{-})$, and $(2S,\,3/2^{+})$, respectively. The $%
(2S,\,1/2^{+})$ state can be assigned either to $\Omega_c(3066)^{0}$ state
or $\Omega_c(3090)^{0}$ excited baryon.
\end{abstract}

\maketitle


\section{Introduction}

\label{sec:Introduction}
The observation by the LHCb Collaboration of new narrow states $\Omega
_{c}^{0}$ in the $\Xi _{c}^{+}K^{-}$ invariant mass distribution is one of
the intriguing discoveries in physics of the heavy baryons \cite{LHCb}.
Preliminary analysis indicates that these five neutral resonances are
composed of $css$ quarks, and may be orbitally/radially excited states of
the $\Omega_c^{0}$ baryons with spins $1/2$ and $3/2$. Let us note, that
till the LHCb data an experimental information about baryons with $css$
content was limited by the masses of the $\Omega _{c}^{0}$ and $\Omega
_{c}(2770)^{0}$ particles \cite{Olive:2016xmw}
\begin{equation}
m=2695.2\pm 1.7\,\mathrm{\,MeV},\,\,m=2765.9\pm 2.0\,\,\mathrm{MeV},
\label{eq:Data}
\end{equation}%
which were considered as the ground states with the spin-parities $%
J^{P}=1/2^{+}$ and $3/2^{+}$, respectively.

Theoretical investigations performed in the context of different approaches,
and predictions obtained for the spectroscopic parameters provide
incomparably more detailed information on the features of the $\Omega_c^{0}$
baryons, than experimental data \cite%
{Capstick:1986bm,Bagan:1991sc,Bagan:1992tp,Chiladze:1997ev,Huang:2000tn,Wang:2002ts, Ebert:2007nw,Ebert:2011kk,Garcilazo:2007eh,Valcarce:2008dr,Roberts:2007ni,Wang:2007sqa, Wang:2008hz,Wang:2009cr,Vijande:2012mk,Yoshida:2015tia,Edwards:2012fx,Padmanath:2013bla, Chen:2015kpa,Chen:2016phw,Shah:2016nxi,Aliev:2015qea,Azizi:2015ksa}%
. In fact, the masses of the ground state and radially/orbitally excited
heavy baryons including the $\Omega_c^{0}$ particles were calculated using
the relativistic quark models \cite%
{Capstick:1986bm,Ebert:2007nw,Ebert:2011kk}, the QCD sum rule method \cite%
{Bagan:1991sc,Bagan:1992tp,Huang:2000tn,Wang:2002ts,
Wang:2007sqa,Wang:2008hz,Wang:2009cr,Chen:2015kpa,Chen:2016phw,Aliev:2015qea,Azizi:2015ksa}%
, the Heavy Quark Effective Theory (HQET) \cite{Chiladze:1997ev}, various
quark models \cite%
{Garcilazo:2007eh,Valcarce:2008dr,Roberts:2007ni,Vijande:2012mk,Yoshida:2015tia,Shah:2016nxi}%
, and lattice simulations \cite{Edwards:2012fx,Padmanath:2013bla}. The
strong couplings and transitions of the heavy flavored baryons, their
magnetic moments and radiative decays also attracted interest of physicists
\cite%
{Aliev:2015qea,Azizi:2015ksa,Zhu:1997as,Aliev:2008sk,Aliev:2009jt,Aliev:2010nh,Aliev:2010ev,Aliev:2011kn,Aliev:2011ufa, Aliev:2011uf,Aliev:2010ac}%
. It is worth noting, that in some of these theoretical studies different
assumptions were made on the structure of the heavy baryons. For example, in
Refs.\ \cite{Ebert:2007nw,Ebert:2011kk} a heavy-quark-light-diquark picture
were employed in the relativistic quark model. In other works, QCD sum rule
calculations were carried out in the context of the HQET \cite%
{Huang:2000tn,Wang:2002ts,Chen:2015kpa,Chen:2016phw}.

The discovery of five new $\Omega_c^{0}$ particles by the LHCb Collaboration
changed the existed experimental situation, and stimulated a theoretical
activity to explain the observed states. These states were seen as
resonances in the $\Xi _{c}^{+}K^{-}$ invariant mass distribution. Their
masses do not differ considerably from each other and are within the range $%
M=3000-3150\, \mathrm{MeV}$. The transition $\Omega _{c}^{0} \to \Xi
_{c}^{+}K^{-}$ may be considered as main decay modes of the $\Omega_c^{0}$
states, widths of which equal to a few $\mathrm{MeV}$.

The LHCb did not provide an information on the spin-parities of the new
states, which is an important problem of ongoing theoretical investigations.
Thus, in our Letter \cite{Agaev:2017jyt} we have calculated the masses of
the ground states and first radial excitations of $\Omega_c^{0}$ with $%
J^{P}=1/2^{+}$ and $3/2^{+}$, and found that the particles $\Omega_c(3066)$
and $\Omega_c(3119)$ can be considered as the radially excited $css$ baryons
with the quantum numbers $(2S,\,1/2^{+})$ and $(2S,\,3/2^{+})$,
respectively. In calculations we have employed the two-point QCD sum rule
method by invoking into analysis general expressions for the currents to
interpolate the $\Omega_c^{0}$ baryons with spins $1/2$ and $3/2$. Our
results correctly describe the masses of the ground states $\Omega_c^{0}$
and $\Omega_c(2770)^{0}$, and agree with two of the recent experimental data
of the LHCb Collaboration. It is interesting, that predictions obtained in
some of previous theoretical studies agree with new LHCb data and our
results (more detailed information can be found in Ref.\ \cite{Agaev:2017jyt},
and in references therein), as well.

The problems connected with the $\Omega_c^{0}$ states have been addressed in
Refs.\ \cite%
{Chen:2017sci,Karliner:2017kfm,Wang:2017vnc,Padmanath:2017lng,Yang:2017rpg,
Huang:2017dwn,Kim:2017jpx,Wang:2017hej,Cheng:2017ove,Wang:2017zjw,Chen:2017,Zhao:2017,Aliev:2017led}%
. The new particles have been assigned to be P-wave $\Omega_c$ baryons in
Ref.\ \cite{Chen:2017sci}, where the authors evaluated widths of their decay
channels. Calculations there have been performed in the framework of HQET
using the sum rule approach. In Refs.\ \cite{Karliner:2017kfm,Wang:2017vnc} $%
\Omega_c(3000)$, $\Omega_c(3050)$, $\Omega_c(3066)$, $\Omega_c(3090)$ and $%
\Omega_c(3119)$ have been interpreted as P-wave excited states of the $%
\Omega_c^{0}$ baryons with the spin-parities $1/2^{-},\,1/2^{-},\,
3/2^{-},\, 3/2^{-}$ and $5/2^{-}$, respectively. In Ref.\ \cite%
{Karliner:2017kfm} an alternative set of assignments, namely $%
3/2^{-},\,3/2^{-},\, 5/2^{-},\, 1/2^{+}$ and $3/2^{+}$ is made for these
states, as well. In this case $1/2^{-} $ states are expected around $2904$ and
$2978 $ $\mathrm{MeV}$. In both of Refs.\ \cite%
{Karliner:2017kfm,Wang:2017vnc} the authors utilized the
heavy-quark-light-diquark model for $\Omega_c$ baryons. On the basis of
lattice simulations the same conclusions have been made also in Ref.\ \cite%
{Padmanath:2017lng}. Attempts have been done to classify new states as
five-quark systems or S-wave pentaquark molecules with $J^{P}=1/2^{-},%
\,3/2^{-}$ and $5/2^{-}$ \cite{Yang:2017rpg,Huang:2017dwn}. The possible
pentaquark interpretation of the $\Omega_c^{0}$ baryons on the basis of the
quark-soliton model has been suggested also in Ref.\ \cite{Kim:2017jpx}.

The explorations carried out in the context of a constituent quark model
have allowed authors of Ref.\ \cite{Wang:2017hej} to conclude, that $\Omega
_{c}(3000)$ and $\Omega _{c}(3090)$ can be considered as states with $%
1/2^{-} $, $\Omega _{c}(3050)$ and $\Omega _{c}(3066)$ as the baryons with $%
3/2^{-}$ and $5/2^{-}$, whereas the $\Omega _{c}(3119)$ might correspond to
one of the radial excitations $(2S,\,1/2^{+})$ or $(2S,\,3/2^{+})$. In Ref.\
\cite{Cheng:2017ove} the first three states from the LHCb range of excited $%
\Omega _{c}$ baryons have been classified as P-wave states with $%
1/2^{-},\,5/2^{-}$ and $3/2^{-}$, whereas last two particles have been
assigned to be $2S$ states with spin-parities $1/2^{+}$ and $3/2^{+}$,
respectively. These states have been analyzed as the P-wave excitation of
the $\Omega _{c}^{0}$ baryons with spin-parities $1/2^{-},\,1/2^{-},%
\,3/2^{-},\,3/2^{-}$ and $5/2^{-}$ also in Ref.\ \cite{Wang:2017zjw}. The
studies have been performed using the two-point sum rule method by
introducing relevant interpolating currents.

The newly discovered $\Omega_c^{0}$ states, their spin-parities has been
analyzed in Refs.\ \cite{Chen:2017,Zhao:2017,Aliev:2017led}, too. Thus,
studies in Ref.\ \cite{Chen:2017} showed that five resonances $\Omega_c^{0}$ can 
be grouped into the 1P states with negative parity, i.e. the resonances  $\Omega_{c}(3000)$ 
and $\Omega _{c}(3090)$ have been considered there as $(1P, 1/2^{-})$ states, $\Omega _{c}(3066)$ and 
$\Omega _{c}(3119)$ as resonances with $(1P,\, 3/2^{-})$, and $\Omega _{c}(3050)$ as 
$(1P,\, 5/2^{-})$ state. The alternative explanation has been suggested in Ref.\  \cite{Zhao:2017}, 
where the resonances $\Omega _{c}(3066)$ and $\Omega _{c}(3090)$ have been interpreted as
$1P$-wave states with the spin-parity $J^{P}=3/2^{-}$ or $J^{P}=5/2^{-}$. 
Starting from decay features of the remaining three resonances in Ref.\  \cite{Zhao:2017} 
the authors have assigned them to be $1D$-wave $\Omega_c^{0}$ states.
Finally, in Ref.\ \cite{Aliev:2017led} the resonances $\Omega_c(3000)$ and $%
\Omega_c(3066)$ have been classified as the $(1P,\ 1/2^{-})$ and $(1P,\
3/2^{-})$ states, respectively.

As is seen, a variety of suggestions made on the structures of the $\Omega
_{c}^{0}$ states, methods and schemes used to compute their parameters, and
obtained predictions for the spin-parities of these baryons is quite
impressive. In the present work we are going to extend our previous paper by
including into analysis P-wave $(1P,\,1/2^{-})$ and $(1P,\,3/2^{-})$ states,
as well. We will evaluate the masses and pole residues of the ground and
four excited $\Omega _{c}^{0}$ states. We will also calculate widths of $%
\Omega _{c}^{0}\rightarrow \Xi ^{+}K^{-}$ decays using the light cone sum
rule (LCSR) method, which is one of the powerful nonperturbative approaches
to evaluate parameters of exclusive processes \cite{Balitsky:1989ry}.
Calculations will be performed by taking into account K meson's distribution
amplitudes (DAs). The extracted from analysis mass and decay width of $%
\Omega _{c}^{0}$ states will be confronted with existing LHCb data, and
predictions obtained in theoretical papers. This will allows us to identify $%
\Omega _{c}(3000)$, $\Omega _{c}(3050)$, $\Omega _{c}(3066)$, and $\Omega
_{c}(3119)$ by fixing their quantum numbers.

This work is structured in the following way. In Sec.\ \ref{sec:mass} we
calculate the mass and pole residue of the ground state and
orbitally/radially excited $\Omega _{c}^{0}$ baryons with the quantum
numbers $(1S,\,1/2^{+})\Rightarrow $ $\Omega _{c}$, $(1P,\,1/2^{-})%
\Rightarrow \Omega _{c}^{-}$, $(2S,\,1/2^{+})$ $\Rightarrow \Omega
_{c}^{\prime }$, and $(1S,\,3/2^{+})\Rightarrow \Omega _{c}^{\star }$, $%
(1P,\,3/2^{-})\Rightarrow \Omega _{c}^{\star -}$, $(2S,\,3/2^{+})\Rightarrow
\Omega _{c}^{\star \prime }$ . To this end, we employ the two-point sum
rules method. In Sec.\ \ref{sec:Decays1/2} we analyze $\Omega _{c}^{-}\Xi
_{c}^{+}K^{-}$ and $\Omega _{c}^{\prime}\Xi _{c}^{+}K^{-}$ vertices to
evaluate the corresponding strong couplings $g_{\Omega ^{-}\Xi K}$ and $%
g_{\Omega ^{\prime }\Xi K}$, and calculate widths of $\Omega _{c}^{-}\to \Xi
_{c}^{+}K^{-}$ and $\Omega _{c}^{\prime }\to \Xi _{c}^{+}K^{-}$ decays. The
similar investigations are carried out in Sec.\ \ref{sec:Decays3/2} for the
vertices containing $\Omega _{c}^{0}$ baryons with $J^{P}=3/2^{+}$ and $%
J^{P}=3/2^{-}$. Here we find widths of the processes $\Omega _{c}^{\star
-}\rightarrow \Xi _{c}^{+}K^{-}$ and $\Omega _{c}^{\star \prime }\rightarrow
\Xi _{c}^{+}K^{-}$. In this section we also analyze the $\Omega _{c}^{\star
\prime }\to \Xi _{c}^{\prime +}K^{-} $ decay, which is kinematically allowed
only for $\Omega _{c}^{\star \prime } $ baryon. Section \ref{sec:Concl} is
reserved for brief discussion of the obtained results. It contains also our
concluding remarks. Explicit expressions of the correlation functions
derived in the present work, as well as the quark propagators used in
calculations are presented in Appendix.


\section{Masses and pole residues of the $\Omega_c^{0}$ states}

\label{sec:mass}

In this section we evaluate the mass and pole residue of the spin $1/2$ and $%
3/2$ ground state and excited $\Omega _{c}$ (hereafter, we omit the
superscript $0$ in $\Omega _{c}^{0}$) baryons by means of two-point sum rule
method.

The sum rules necessary to find masses and residues of the $\Omega _{c}^{0}$
baryons can be derived using the two-point correlation function
\begin{equation}
\Pi _{(\mu \nu )}(p)=i\int d^{4}xe^{ipx}\langle 0|\mathcal{T}\{\eta _{(\mu
)}(x)\overline{\eta }_{(\nu )}(0)\}|0\rangle ,  \label{eq:CorrF1}
\end{equation}%
where $\eta (x)$ and $\eta _{\mu }(x)$ are the interpolating currents for $%
\Omega _{c}$ states with spins $J=1/2$ and $J=3/2$, respectively. They have
the following forms
\begin{eqnarray}
\eta &=&-\frac{1}{2}\epsilon ^{abc}\left\{ \left( s^{aT}Cc^{b}\right) \gamma
_{5}s^{c}+\beta \left( s^{aT}C\gamma _{5}c^{b}\right) s^{c}\right.  \notag \\
&&\left. -\left[ \left( c^{aT}Cs^{b}\right) \gamma _{5}s^{c}+\beta \left(
c^{aT}C\gamma _{5}s^{b}\right) s^{c}\right] \right\} ,  \label{eq:BayC1/2}
\end{eqnarray}%
and
\begin{eqnarray}
\eta _{\mu } &=&\frac{1}{\sqrt{3}}\epsilon ^{abc}\left\{ \left(
s^{aT}C\gamma _{\mu }s^{b}\right) c^{c}+\left( s^{aT}C\gamma _{\mu
}c^{b}\right) s^{c}\right.  \notag \\
&&\left. +\left( c^{aT}C\gamma _{\mu }s^{b}\right) s^{c}\right\} .
\label{eq:BayC3/2}
\end{eqnarray}%
In expressions above $C$ is the charge conjugation operator. The current $%
\eta (x)$ \ for the $1/2$ baryons contains an arbitrary auxiliary parameter $%
\beta $, where $\beta =-1$ corresponds to the Ioffe current.

We start from the spin $1/2$ baryons and calculate the correlation function $%
\Pi ^{\mathrm{Phys}}(p)$ in terms of the physical parameters of the states
under consideration and determine $\Pi ^{\mathrm{OPE}}(p)$ employing the
quark propagators. Because, the current $\eta (x)$ couples not only to
states $\Omega _{c}$ and $\Omega _{c}^{\prime }$, but also to $\Omega
_{c}^{-}$, in the physical side of the sum rule we explicitly take into
account their contributions by adopting the "ground-state+first
orbitally+first radially excited states+continuum" scheme: We follow an
approach applied recently to calculate the masses and residues of radially
excited octet and decuplet baryons in Refs. \cite%
{Aliev:2016jnp,Aliev:2016adl}. In these works the authors got results, which
are compatible with existing experimental data on the masses of the radially
excited baryons, and demonstrated that besides ground-state baryons the QCD
sum rule method can be successfully applied to investigate their
excitations, as well.

Thus, we find
\begin{eqnarray}
&&\Pi ^{\mathrm{Phys}}(p)=\frac{\langle 0|\eta |\Omega _{c}(p,s)\rangle
\langle \Omega _{c}(p,s)|\overline{\eta }|0\rangle }{m^{2}-p^{2}} \notag \\
&&+\frac{\langle 0|\eta |\Omega _{c}^{-}(p,\widetilde{s})\rangle \langle \Omega
_{c}^{-}(p,\widetilde{s})|\overline{\eta }||0\rangle }{\widetilde{m}%
^{2}-p^{2}}  \notag \\
&&+\frac{\langle 0|\eta |\Omega _{c}^{\prime }(p,s^{\prime })\rangle \langle
\Omega _{c}^{\prime }(p,s^{\prime })|\overline{\eta }||0\rangle }{m^{\prime
2}-p^{2}}+\ldots ,  \label{eq:CF1/2}
\end{eqnarray}%
where  $m$, $\widetilde{m}$, $m^{\prime }$ and $s$, $%
\widetilde{s}$, $s^{\prime }$ are the masses and spins of the $\Omega _{c}$,
$\Omega _{c}^{-}$ and $\Omega _{c}^{\prime }$ baryons, respectively. The dots denote
contributions of higher resonances and continuum states. In Eq.\ (\ref{eq:CF1/2})
the summations over the spins $s$, $\widetilde{s}$, $s^{\prime }$ are implied.

We proceed by introducing the matrix elements
\begin{eqnarray}
&&\langle 0|\eta |\Omega _{c}^{(\prime )}(p,s^{(\prime )})\rangle =\lambda
^{(\prime )}u^{(\prime )}( p,s^{(\prime )}) ,  \notag \\
&&\langle 0|\eta |\Omega _{c}^{-}(p,\widetilde{s})\rangle =\widetilde{%
\lambda }\gamma _{5}u^{-}(p,\widetilde{s}).  \label{eq:MElem}
\end{eqnarray}%
Here $\lambda $, $\widetilde{\lambda }$ and $\lambda ^{\prime }$ are the
pole residues of the $\Omega _{c}$, $\Omega _{c}^{-}$ and $\Omega
_{c}^{\prime }$ states, respectively. Using Eqs.\ (\ref{eq:CF1/2}) and (\ref%
{eq:MElem}) and carrying out summation over spins of the $1/2$ baryons
\begin{equation}
\sum\limits_{s}u(p,s)\overline{u}(p,s)=\slashed p+m,
\end{equation}%
we obtain%
\begin{eqnarray}
&&\Pi ^{\mathrm{Phys}}(p)=\frac{\lambda ^{2}(\slashed p+m)}{m^{2}-p^{2}}+%
\frac{\widetilde{\lambda }^{2}(\slashed p-\widetilde{m})}{\widetilde{m}%
^{2}-p^{2}}  \notag \\
&&+\frac{\lambda ^{\prime 2}(\slashed p+m^{\prime })}{m^{\prime 2}-p^{2}}%
+\ldots   \label{eq:CorrF2}
\end{eqnarray}%
The Borel transformation of this expression is:
\begin{eqnarray}
&&\mathcal{B}\Pi ^{\mathrm{Phys}}(p)=\lambda ^{2}e^{-\frac{m^{2}}{M^{2}}}(%
\slashed p+m)  \notag \\
&&+\widetilde{\lambda }^{2}e^{-\frac{\widetilde{m}^{2}}{M^{2}}}(\slashed p-%
\widetilde{m})+\lambda ^{\prime 2}e^{-\frac{m^{\prime 2}}{M^{2}}}(\slashed %
p+m^{\prime }).  \label{eq:Bor1}
\end{eqnarray}%
As is seen, it contains the structures $\sim \slashed p$ and $\sim I$. In
order to derive the sum rules we use both of them and find: from the terms $%
\sim \slashed p$
\begin{equation}
\lambda ^{2}e^{-\frac{m^{2}}{M^{2}}}+\widetilde{\lambda }^{2}e^{-\frac{%
\widetilde{m}^{2}}{M^{2}}}+\lambda ^{\prime 2}e^{-\frac{m^{\prime 2}}{M^{2}}%
}=\mathcal{B}\Pi _{1}^{\mathrm{OPE}}(p),  \label{eq:MFor1}
\end{equation}%
and from the terms $\sim I$
\begin{equation}
\lambda ^{2}me^{-\frac{m^{2}}{M^{2}}}-\widetilde{\lambda }^{2}\widetilde{m}%
e^{-\frac{\widetilde{m}^{2}}{M^{2}}}+\lambda ^{\prime 2}m^{\prime }e^{-\frac{%
m^{\prime 2}}{M^{2}}}=\mathcal{B}\Pi _{2}^{\mathrm{OPE}}(p),
\label{eq:MFor2}
\end{equation}%
where $\mathcal{B}\Pi _{1}^{\mathrm{OPE}}(p)$ and $\mathcal{B}\Pi _{2}^{%
\mathrm{OPE}}(p)$ are the Borel transformations of the same structures in $%
\Pi ^{\mathrm{OPE}}(p)$ computed employing the quark propagators, as it has
been explained above. It is assumed, that continuum contributions are
subtracted from the right-hand sides of Eqs.\ (\ref{eq:MFor1}) and (\ref%
{eq:MFor2}) utilizing the quark-hadron duality assumption.

The derived sum rules contain six unknown parameters of the ground state and
excited baryons. Therefore, from Eqs.\ (\ref{eq:MFor1}) and (\ref{eq:MFor2})
we determine the parameters $(m,\ \lambda )$ of the ground state $\Omega
_{c} $ baryon by keeping there only the first terms, and choosing
accordingly the continuum threshold parameter $s_{0}$ in $\Pi _{1}^{\mathrm{%
OPE}}(M^2,\ s_{0})$ and $\Pi_{2}^{\mathrm{OPE}}(M^2,\ s_{0})$: This is sum
rule's computations within the "ground-state + continuum" scheme. At the
next step, we retain in the sum rules terms corresponding to $\Omega _{c}$
and $\Omega _{c}^{-}$ baryons, but treat $(m,\ \lambda )$ as input
parameters to extract $(\widetilde{m},\ \widetilde{\lambda })$: the
continuum threshold now is chosen as $\widetilde{s}_{0}>s_{0}$. Finally, the
set of $(m,\ \lambda )$ and $(\widetilde{m},\ \widetilde{\lambda })$ is
utilized in the full version of the sum rules to find parameters $(m^{\prime
},\ \lambda ^{\prime })$ of the $\Omega _{c}^{\prime }$ baryon, with $%
s_{0}^{\prime }>\widetilde{s}_{0}$ being the relevant continuum threshold.

The similar analysis with additional technical details is valid also for the
spin $3/2$ baryons, as well. Indeed, in this case we use the matrix elements
\begin{eqnarray}
&&\langle 0|\eta _{\mu }|\Omega _{c}^{\ast (\prime )}(p,s^{(\prime
)})\rangle =\lambda ^{(\prime )}u_{\mu }^{(\prime )}(p,s^{(\prime )}),
\notag \\
&&\langle 0|\eta _{\mu }|\Omega _{c}^{\ast -}(p,\widetilde{s})\rangle =%
\widetilde{\lambda }\gamma _{5}u_{\mu }^{-}(p,\widetilde{s})
\end{eqnarray}%
where $u_{\mu }(p,s)$ are the Rarita-Schwinger spinors, and carry out the
summation over $s$ by means of the formula
\begin{eqnarray}
&&\sum\limits_{s}u_{\mu }(p,s)\overline{u}_{\nu }(p,s)=-(\slashed p+m)\left[
g_{\mu \nu }-\frac{1}{3}\gamma _{\mu }\gamma _{\nu }\right.   \notag \\
&&\left. -\frac{2}{3m^{2}}p_{\mu }p_{\nu }+\frac{1}{3m}(p_{\mu }\gamma _{\nu
}-p_{\nu }\gamma _{\mu })\right] .
\end{eqnarray}%
The interpolating current $\eta _{\mu }$ couples to spin-$1/2$ baryons,
therefore the sum rules contain contributions arising from these terms.
Their undesired effects can be eliminated by applying a special ordering of
the Dirac matrices (see, for example Ref.\ \cite{Aliev:2016jnp}). It is not
difficult to demonstrate, that structures $\sim \slashed pg_{\mu \nu }$ and $%
\sim g_{\mu \nu }$ are free of contaminations and formed only due to
contributions of spin-$3/2$ baryons. In order to derive the sum rules for
the masses and pole residues of the ground-state and excited $\Omega _{c}^{0}
$ baryons with spin-parities $3/2^{-}$ and $3/2^{+}$ we employ only these
structures and corresponding invariant amplitudes.

The correlation functions $\Pi(p)$ and $\Pi_{\mu\nu}(p)$ should be found
using the quark propagators: This is necessary to get the QCD side of the
sum rules. We calculate them employing the general expression given by Eq.\ (%
\ref{eq:CorrF1}) and currents defined in Eqs.\ (\ref{eq:BayC1/2}) and (\ref%
{eq:BayC3/2}). The results for $\Pi^{\mathrm{OPE}}(p)$ and $\Pi_{\mu\nu}^{%
\mathrm{OPE}}(p)$ in terms of the $s$ and $c$-quarks' propagators are
written down in Appendix. Here we also present analytic expressions of the
propagators, themselves. Manipulations to calculate correlators using
propagators in the coordinate representation, to extract relevant two-point
spectral densities and perform the continuum subtraction are well known and were
numerously described in the existing literature. Therefore, we do not
 concentrate further on details of these rather lengthy computations.

The sum rules contain the vacuum expectations values of the different
operators and masses of the $s$ and $c$-quarks, which are input parameters
in the numerical calculations. The vacuum condensates are well known: for
the quark and mixed condensates we use $\langle \overline{s}s\rangle
=-0.8\cdot (0.24\pm 0.01)^{3}~\mathrm{GeV}^{3}$, $\langle \overline{s}%
g_{s}\sigma Gs\rangle =m_{0}^{2}\langle \overline{s}s\rangle $, where $%
m_{0}^{2}=(0.8\pm 0.1)~\mathrm{GeV}^{2}$, whereas for the gluon condensate
we utilize $\langle \alpha _{s}G^{2}/\pi \rangle =(0.012\pm 0.004)~\mathrm{%
GeV}^{4}$. The masses of the strange and charmed quarks are chosen equal to $%
m_{s}=96_{-4}^{+8}\ \mathrm{MeV}$ and $m_{c}=(1.27\pm 0.03)\ \mathrm{GeV}$,
respectively. These parameters and their different products determine an
accuracy of performed numerical computations: In the present work we take
into account terms up to ten dimensions.

The sum rules depend also on the auxiliary parameters $M^2$ and $s_0$, which
are not arbitrary, but can be changed within special regions. Inside of
these working regions the convergence of the operator product expansion,
dominance of the pole contribution over remaining terms should be satisfied.
The prevalence of the perturbative contribution in the sum rules, and
relative stability of the extracted results are also among restrictions of
calculations. At the same time, the Borel and continuum threshold parameters
are main sources of ambiguities, which affect final predictions
considerably. These uncertainties may amount to $ 30\%$ of results, and are
unavoidable features of sum rules' predictions. For spin-$1/2$ particles
there is an additional dependence on $\beta$, stemming from the expression
of the interpolating current $\eta(x)$. The choice of an interval for $\beta$
should also obey the clear requirement: we fix the working region for $\beta$
by demanding a weak dependence of our results on its choice. Results for the
spin-$1/2$ particles are obtained by varying $\beta =\tan \theta $ within
the limits
\begin{equation}
-0.75\leq \cos \theta \leq -0.45,\ 0.45\leq \cos \theta \leq 0.75,
\label{eq:Beta}
\end{equation}%
where we have achieved best stability of our predictions. Let us note, that
for the famous Ioffe current $\cos\theta=-0.71$.
\begin{table}[tbp]
\begin{tabular}{|c|c|c|c|c|}
\hline
$(n,J^{P})$ & $(1S,\frac{1}{2}^{+})$ & $(1P,\frac{1}{2}^{-})$ & $(2S,\frac{1%
}{2}^{+})$ &  \\ \hline
$M^2 ~(\mathrm{GeV}^2$) & $3.5-5.5$ & $3.5-5.5$ & $3.5-5.5$ &  \\ \hline
$s_0 ~(\mathrm{GeV}^2$) & $3.0^2-3.2^2$ & $3.3^2-3.5^2$ & $3.5^2-3.7^2$ &
\\ \hline
$m_{\Omega_c} ~(\mathrm{MeV})$ & $2685\pm 123$ & $2990 \pm 129$ & $3075 \pm
142$ &  \\ \hline
$\lambda_{\Omega_c}\cdot 10^{2} ~(\mathrm{GeV}^3)$ & $6.2 \pm 1.8$ & $11.9
\pm 2.8$ & $17.1 \pm 3.4$ &  \\ \hline
\end{tabular}%
\caption{The sum rule results for the masses and residues of the $%
\Omega_c^{0}$ baryons with the spin-$1/2$.}
\label{tab:Results1A}
\end{table}
\begin{table}[tbp]
\begin{tabular}{|c|c|c|c|c|}
\hline
$(n,J^{P})$ & $(1S,\frac{3}{2}^{+})$ & $(1P,\frac{3}{2}^{-})$ & $(2S,\frac{3%
}{2}^{+})$ &  \\ \hline
$M^2 ~(\mathrm{GeV}^2$) & $3.5-5.5$ & $3.5-5.5$ & $3.5-5.5$ &  \\ \hline
$s_0 ~(\mathrm{GeV}^2$) & $3.1^2-3.3^2$ & $3.4^2-3.6^2$ & $3.6^2-3.8^2$ &
\\ \hline
$m_{\Omega_c} ~(\mathrm{MeV})$ & $2769\pm 89$ & $3056 \pm 103$ & $3119 \pm
108$ &  \\ \hline
$\lambda_{\Omega_c}\cdot 10^{2} ~(\mathrm{GeV}^3)$ & $7.1 \pm 1.0$ & $16.1
\pm 1.8$ & $25.0 \pm 3.1$ &  \\ \hline
\end{tabular}%
\caption{The predictions for the masses and residues of the spin $3/2$ $%
\Omega_c^{0}$ baryons.}
\label{tab:Results2A}
\end{table}

Results obtained in this work for the masses and residues of the spin-$1/2$
and $3/2$ $\Omega_c$ baryons are presented in Tables\ \ref{tab:Results1A}
and \ref{tab:Results2A}, respectively. Here we also provide the working
windows for parameters $M^2$ and $s_0$ used in extracting $m$ and $\lambda$.
The masses and pole residues of the radially excited baryons $(2S,\ 1/2^{+})$
and $(2S,\ 3/2^{+})$ slightly differ from predictions obtained for these
states in our previous work \cite{Agaev:2017jyt}. These unessential
differences can be explained by features of schemes adopted in Ref.\ \cite%
{Agaev:2017jyt} and in the present work. In fact, in Ref.\ \cite%
{Agaev:2017jyt} parameters of the radially excited states were extracted
within the "ground-state+2S-state+continuum" approximation, whereas now we
apply the "ground-state+1P+2S-states+continuum" scheme: An additional baryon
included into analysis, naturally affects final predictions.

In order to explore a sensitivity of the obtained results on the Borel parameter $M^2$
and continuum threshold $s_0$, in Figs.\ \ref{fig:Mass1}, \ref{fig:Mass2} and \ref{fig:Mass3}
we depict the  $\Omega_c^{\star}$,  $\Omega_c^{\star -}$ and
$\Omega_c^{\star \prime}$ baryons' masses as
functions of these parameters. It is seen, that the dependence of the masses on the parameters
$M^2$ and $s_0$  is mild. In Fig.\ \ref{fig:Residue1} we show, as an example,  the dependence
of the ground-state $\Omega_c^{\star}$ baryon's
residue on the auxiliary parameters of the sum rule computations. The observed behavior
of $\lambda$ on $M^2$ and $s_0$ is typical for such kind of quantities: The
systematic errors  are within limits accepted in the sum rule method.
The sum rule predictions for the masses and residues of the spin-1/2 baryons $\Omega_c$,
$\Omega_c^{-}$ and $\Omega_c^{\prime}$ demonstrate the similar dependence on the
Borel parameter $M^2$ and continuum threshold $s_0$, therefore we refrain from providing
corresponding graphics here.
\begin{widetext}

\begin{figure}[h!]
\begin{center}
\includegraphics[totalheight=6cm,width=8cm]{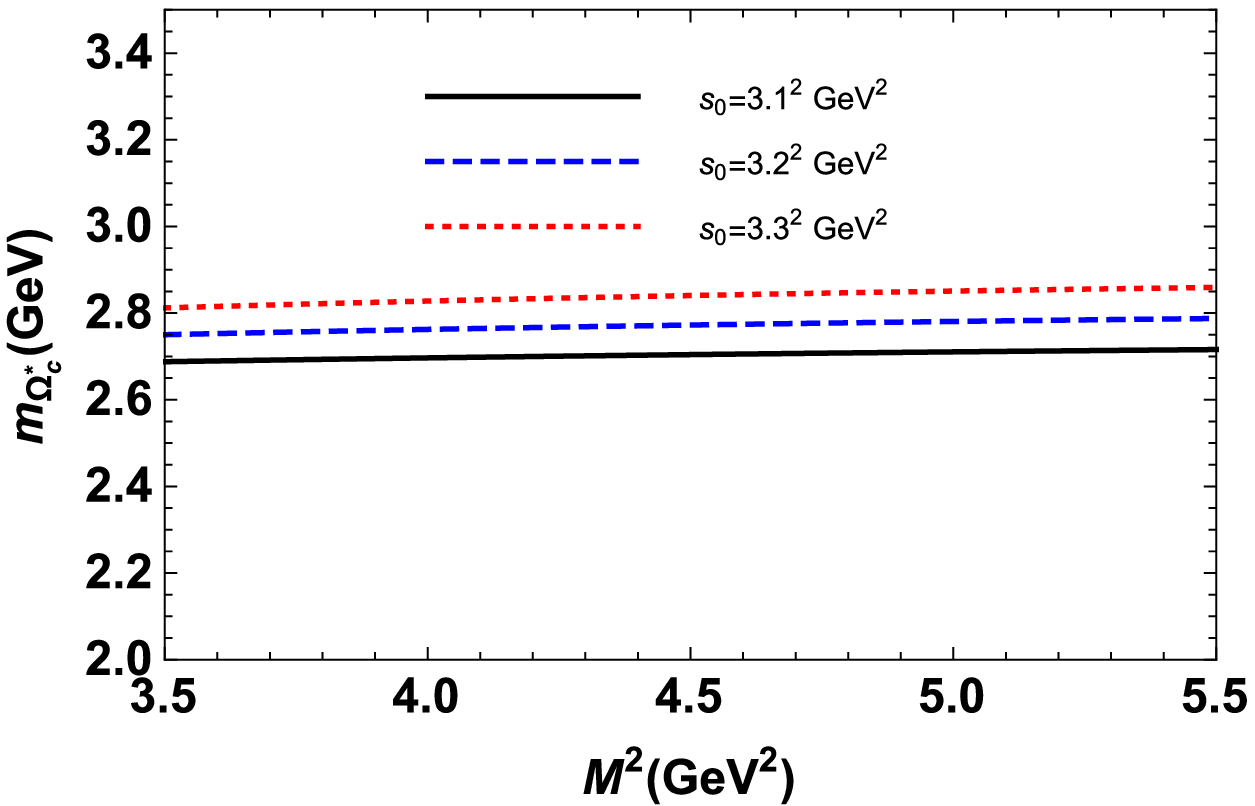}\,\, %
\includegraphics[totalheight=6cm,width=8cm]{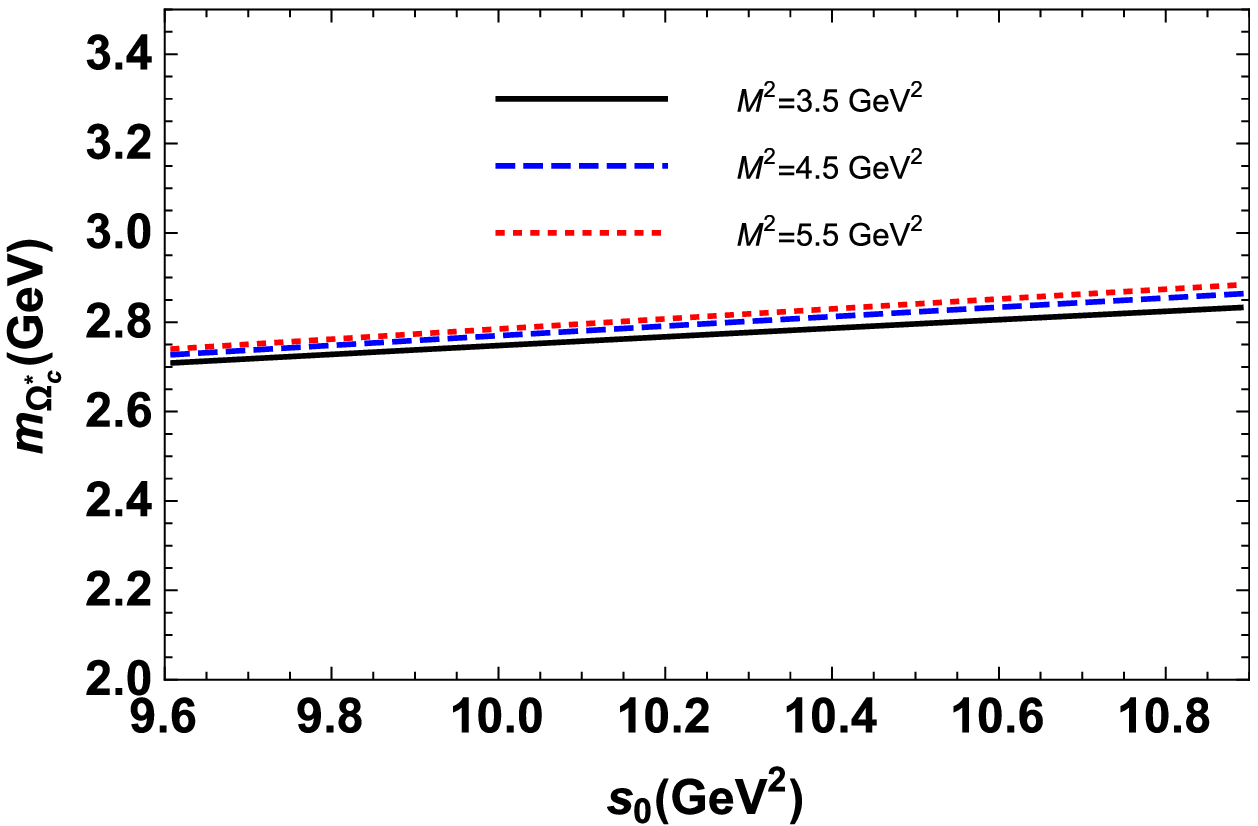}
\end{center}
\caption{ The mass of the ground-state  $\Omega_c^{\star}$ baryon as a function of the Borel parameter $M^2$ at fixed $s_0$ (left panel), and as a function of the continuum threshold $s_0$ at fixed $M^2$ (right panel).}
\label{fig:Mass1}
\end{figure}

\begin{figure}[h!]
\begin{center}
\includegraphics[totalheight=6cm,width=8cm]{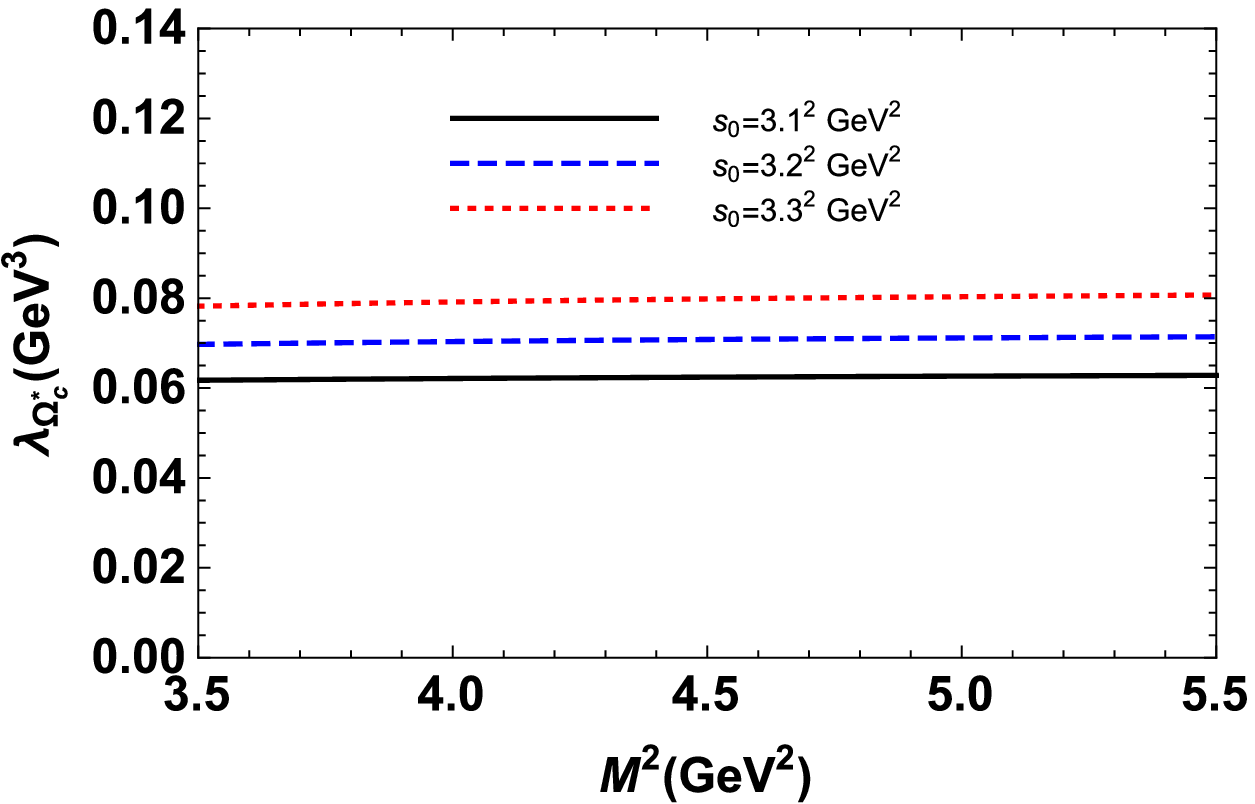}\,\, %
\includegraphics[totalheight=6cm,width=8cm]{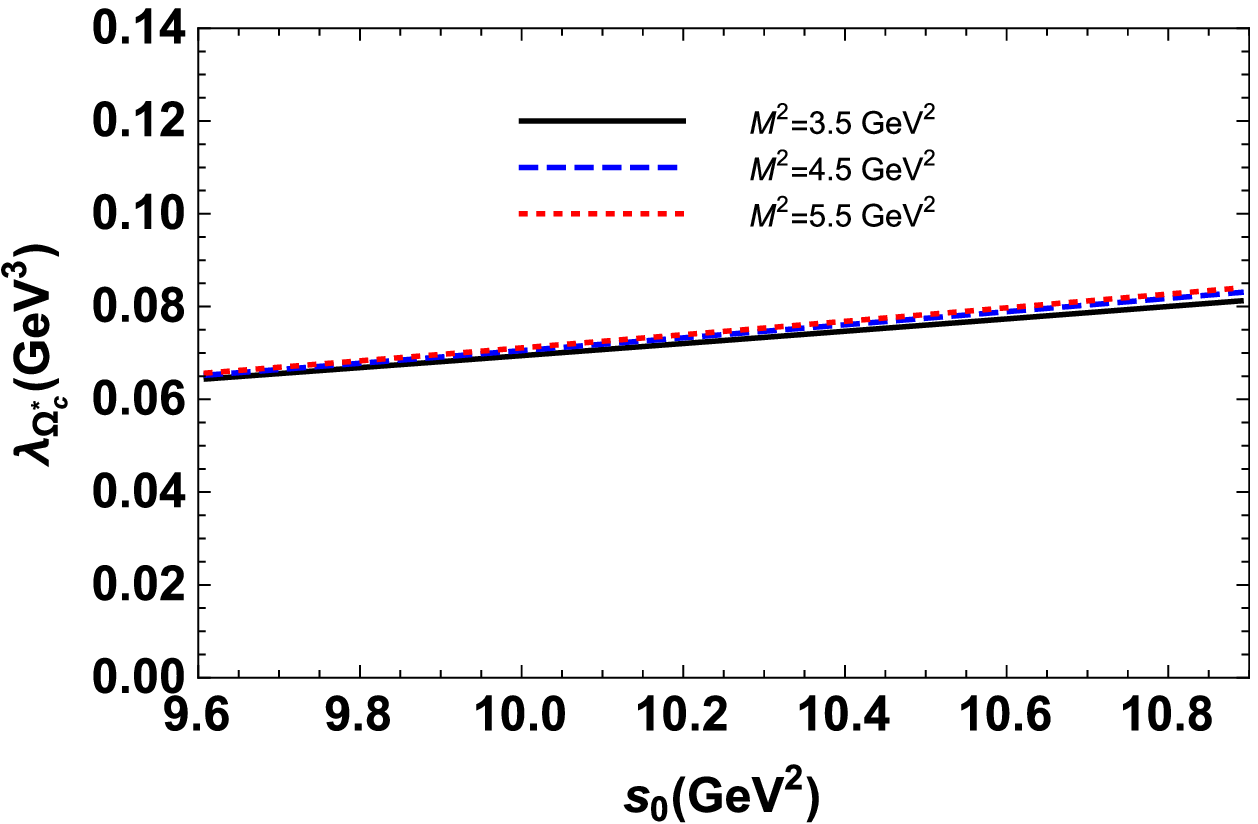}
\end{center}
\caption{ The dependence of the $\Omega_c^{\star}$ baryon's residue $\lambda_{\Omega_c^{\star}}$ on the Borel parameter $M^2$ at chosen values of $s_0$ (left panel), and on the $s_0$ at fixed $M^2$  (right panel).}
\label{fig:Residue1}
\end{figure}

\begin{figure}[h!]
\begin{center}
\includegraphics[totalheight=6cm,width=8cm]{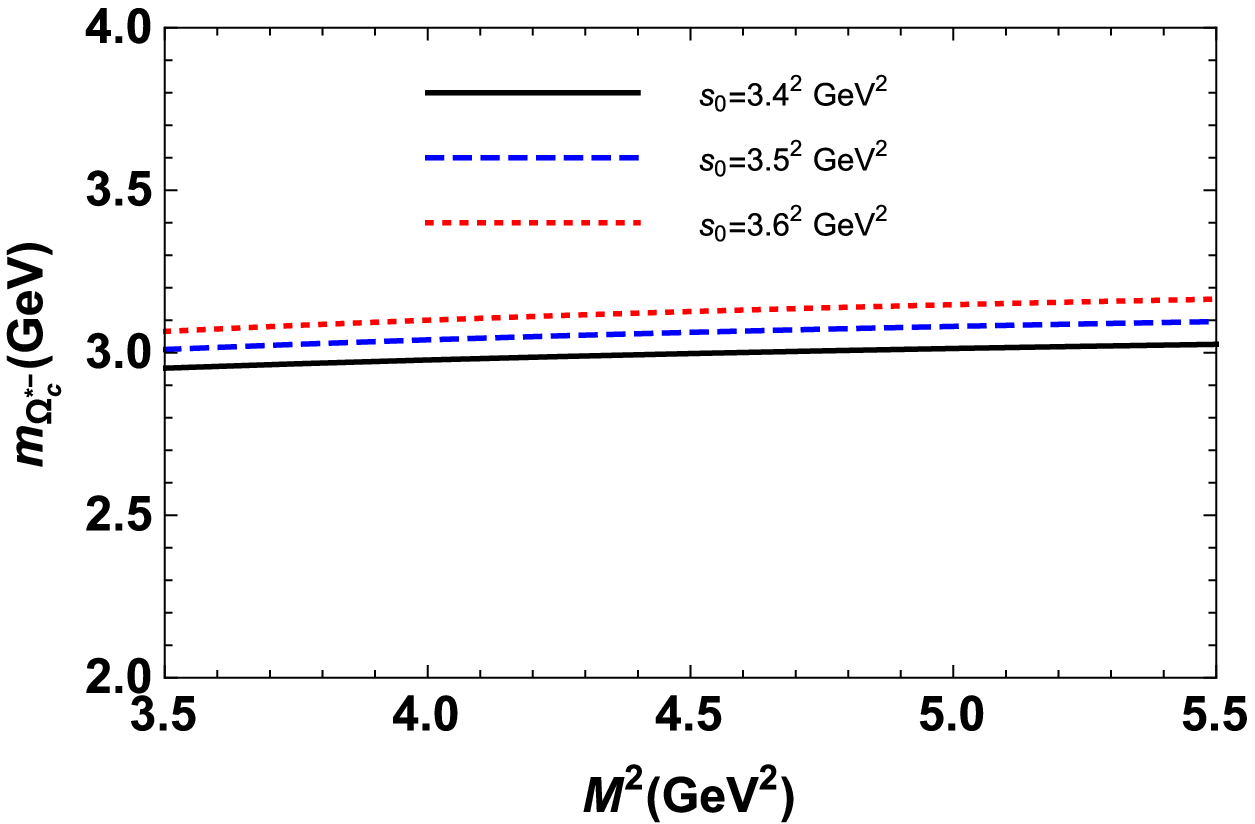}\,\, %
\includegraphics[totalheight=6cm,width=8cm]{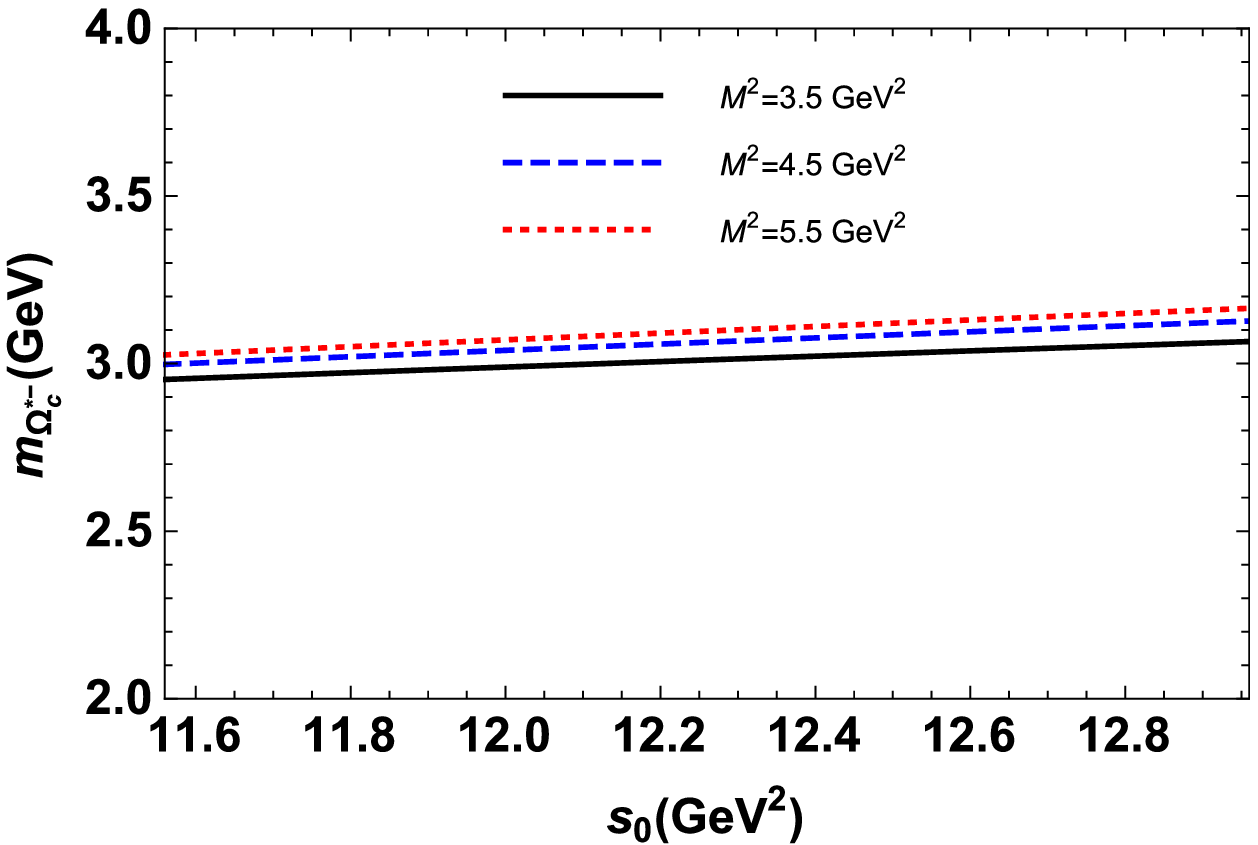}
\end{center}
\caption{ The same as in Fig.\ \ref{fig:Mass1}, but for the orbitally  excited $\Omega_c^{\star -}$ baryon.}
\label{fig:Mass2}
\end{figure}

\begin{figure}[h!]
\begin{center}
\includegraphics[totalheight=6cm,width=8cm]{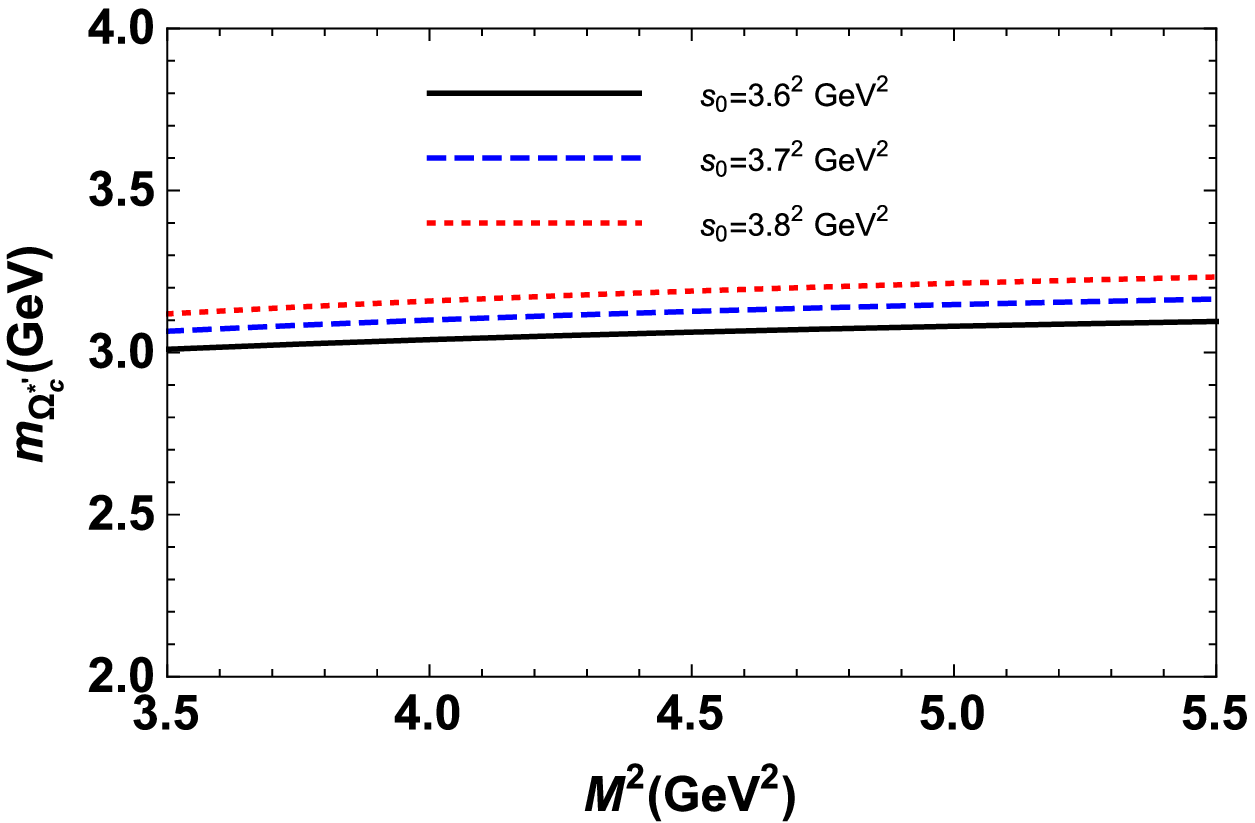}\,\, %
\includegraphics[totalheight=6cm,width=8cm]{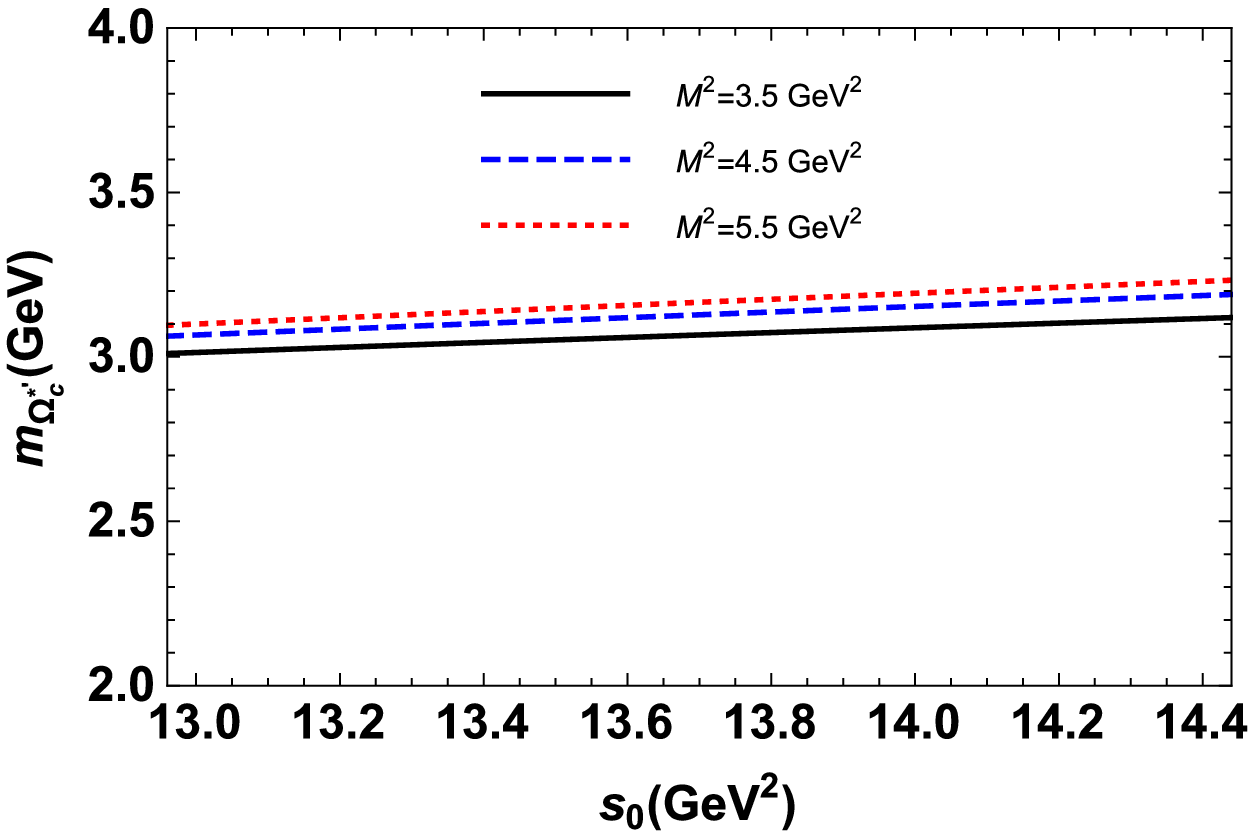}
\end{center}
\caption{ The same as in Fig.\ \ref{fig:Mass1}, but for the radially excited $\Omega_c^{\star \prime}$ baryon. }
\label{fig:Mass3}
\end{figure}

\end{widetext}

It is instructive to explore the "convergence" of the iterative process used
in the present work to evaluate parameters of the $\Omega_c$ baryons. It is known,
that the ground-state contributes dominantly to the spectral density.  The excited
states included into the sum rules are sub-leading terms. To quantify this statement
we  calculate the pole contribution (PC) to the sum rules in the successive stages
of the iterative process to reveal effects due to the ground-state and excited baryons.
To this end, we fix  the Borel parameter $M^2=4.5\ \mathrm{GeV}^2$ (for spin-1/2 baryons also
$\cos \theta = -0.5$) and compute the PC at each stage using for the continuum threshold
$s_0$ its upper limit from the relevant intervals (see, Tables\ \ref{tab:Results1A} and
\ref{tab:Results2A}). We start from the spin-1/2 baryons and from the "ground-state + continuum"
phase, and find that PC arising from $\Omega_c$ equals to $44 \%$ of the result.
Computations in the  "ground-state + 1P state + continuum" step allows us to fix the total
PC from $\Omega_c$ and $\Omega_c^{-}$ baryons at the level $58 \%$ of the whole
prediction, or $14 \%$ effect appearing due to $\Omega_c^{-}$. Finally, in the
"ground-state + 1P + 2S states + continuum" stage the  PC arising from the
$\Omega_c$, $\Omega_c^{-}$ and  $\Omega_c^{\prime}$ baryons amounts to $68 \%$ of the sum
rules, which indicates $10 \%$ contribution of the $\Omega_c^{\prime}$ baryon. The same
analysis carried out for the spin-3/2 baryons leads to the following results:
the ground-state baryon $\Omega_c^{\star}$ forms $41 \%$ of the sum rule, whereas the
excited states $\Omega_c^{\star -}$ and $\Omega_c^{\star \prime}$ constitute $15 \%$ and  $9 \%$
of the whole prediction, respectively. It is worth noting that dependence 
of the presented estimations  on $M^2$ and $\cos \theta$ is negligible.

It is seen, that the procedure adopted in the present work is consistent with general
principles of the sum rule calculations. Because contributions of the higher excited
states decrease, it is legitimate to restrict analysis by considering only two of them. But there are
another reasons to truncate the iterative process at this phase. Indeed, the next spin-1/2 excited baryons
in this range should be $(2P, 1/2^{-})$ and $(3S, 1/2^{+})$  states. By taking into account the
 mass splitting between $\Omega_c$ and first orbitally and radially excited
$\Omega_c^{-}$ and  $\Omega_c^{\prime}$ baryons, it is not difficult to anticipate that
masses of the $(2P, 1/2^{-})$ and $(3S, 1/2^{+})$ states will be higher than recent LHCb data.
The same arguments are valid for the spin-3/2 baryons, as well. The parameters of the higher excited
states of  $\Omega_c$ and $\Omega_c^{\star}$ baryons may provide a valuable information on their
properties, which are interesting for hadron spectroscopy, nevertheless, this task is
beyond the scope of the present investigation.

Basing on the results for the masses of $\Omega _{c}^{0}$ baryons, taking
into account the central values in the sum rules' predictions, and comparing
them with LHCb data we, at this stage of our investigations, assign the
orbitally and radially excited $\Omega _{c}^{0}$ baryons to be newly
discovered states, as it is shown in Table\ \ref{tab:Lit1}. Thus, we have
correlated the excited $\Omega _{c}^{0}$ baryons to states, which were
recently observed by the LHCb Collaboration. Nevertheless, we consider this
assignment as a preliminary one, because systematic errors in sum rule
calculations are significant, and robust conclusions can be drawn only after
analysis of the width of decays $\Omega _{c}^{0}\rightarrow \Xi
_{c}^{+}K^{-} $ and $\Omega _{c}^{0}\rightarrow \Xi _{c}^{\prime +}K^{-}$.

\begin{widetext}

\begin{table}[tbp]
\begin{tabular}{|c|c|c|c|c|c|c|c|}
\hline\hline
& $\Omega_c^{0}$ & $\Omega_c(2770)^{0}$ &  $\Omega_c(3000)^{0}$ &
$\Omega_c(3050)^{0}$ & $\Omega_c(3066)^{0}$ & $\Omega_c(3090)^{0}$ & $\Omega_c(3119)^{0}$  \\
& ($\mathrm MeV$) & ($\mathrm MeV$) & ($\mathrm MeV$) & ($\mathrm MeV$) & ($\mathrm MeV$) & ($\mathrm MeV$) & ($\mathrm MeV$) \\ \hline
$(n,J^{P})$& $(1S,\frac{1}{2}^{+})$ & $(1S,\frac{3}{2}^{+})$ & $(1P,\frac{1}{2}^{-})$ & $(1P,\frac{3}{2}^{-})$ & $(2S,\frac{1%
}{2}^{+})$ & $-$ & $(2S,\frac{3}{2}^{+})$ \\ \hline\hline
Ref.\ \cite{LHCb} & $-$ & $-$ & $3000.4 \pm 0.2 \pm 0.1$ & $3050.2 \pm 0.1 \pm 0.1 $ & $3065.6 \pm 0.1 \pm 0.3$ & $3090.2 \pm 0.3 \pm 0.5$ & $ 3119.1 \pm 0.3 \pm 0.9 $ \\ \hline
Ref.\ \cite{Olive:2016xmw}& $2695.2 \pm 1.7$ & $2765.9 \pm 2.0$ & $-$ & $-$ & $-$ & $-$ & $-$ \\ \hline
This w. & $2685 \pm 123$ & $2769 \pm 89$  & $2990 \pm 129$  & $3056 \pm 103$ & $3075 \pm 142$ & $-$ & $3119 \pm 108$ \\
\hline\hline
\end{tabular}%
\caption{Our results for the masses of  $\Omega_c^{0}$ baryons with spins  $1/2$ and $3/2$, and experimental data from Refs.\ \cite{LHCb,Olive:2016xmw}.}
\label{tab:Lit1}
\end{table}

\end{widetext}


\section{$\Omega _{c}^{-}$ and $\Omega _{c}^{\prime }$ transitions to $%
\Xi_{c}^{+}K^{-}$}

\label{sec:Decays1/2}
The results for the masses of the excited $\Omega_c^{0}$ baryons show that
all of them are above the $\Xi_c^{+}K^{-}$ threshold. Hence, these four
states can decay through the $\Omega _{c}^{0} \to \Xi _{c}^{+}K^{-}$
channels.

In this section we study the vertices $\Omega _{c}^{-}\Xi _{c}^{+}K^{-}$ and
$\Omega _{c}^{\prime }\Xi _{c}^{+}K^{-}$, and calculate corresponding strong
couplings $g_{\Omega ^{-}\Xi K}$ and $g_{\Omega ^{\prime }\Xi K}$ (the
sub-index $c $ is omitted from the baryons for simplicity), which are
necessary to calculate widths of the decays $\Omega _{c}^{-}\rightarrow \Xi
_{c}^{+}K^{-}$ and $\Omega _{c}^{\prime }\rightarrow \Xi _{c}^{+}K^{-}$. To
this end we introduce the correlation function%
\begin{equation}
\Pi (p,q)=i\int d^{4}xe^{ipx}\langle K(q)|\mathcal{T}\{\eta _{\Xi _{c}}(x)%
\overline{\eta }(0)\}|0\rangle ,
\end{equation}%
where $\eta _{\Xi _{c}}(x)$ is the interpolating current for the $\Xi
_{c}^{+}$ baryon. The $\Xi _{c}^{+}$ belongs to the anti-triplet
configuration of the heavy baryons with a single heavy quark. Its current is
anti-symmetric with respect to exchange of the two light quarks, and has the
form
\begin{eqnarray}
\eta _{\Xi _{c}} &=&\frac{1}{\sqrt{6}}\epsilon ^{abc}\left\{ 2\left(
u^{aT}Cs^{b}\right) \gamma _{5}c^{c}+2\beta \left( u^{aT}C\gamma
_{5}s^{b}\right) c^{c}\right.  \notag \\
&&+\left( u^{aT}Cc^{b}\right) \gamma _{5}s^{c}+\beta \left( u^{aT}C\gamma
_{5}c^{b}\right) s^{c}  \notag \\
&&\left. +\left( c^{aT}Cs^{b}\right) \gamma _{5}u^{c}+\beta \left(
c^{aT}C\gamma _{5}s^{b}\right) u^{c}\right\} .  \label{eq:XiCurr}
\end{eqnarray}

We first represent the correlation function $\Pi (p,q)$ using the parameters
of the involved baryons, and, by this manner determine the phenomenological
side of the sum rules. We get%
\begin{eqnarray}
&&\Pi ^{\mathrm{Phys}}(p,q)=\frac{\langle 0|\eta _{\Xi _{c}}|\Xi
_{c}(p,s)\rangle }{p^{2}-m_{\Xi _{c}}^{2}}\langle K(q)\Xi _{c}(p,s)|\Omega
_{c}^{-}(p^{\prime },s^{\prime })\rangle  \notag \\
&&\times \frac{\langle \Omega _{c}^{-}(p^{\prime },s^{\prime })|\overline{%
\eta }|0\rangle }{p^{\prime 2}-\widetilde{m}^{2}}+\frac{\langle 0|\eta _{\Xi
_{c}}|\Xi _{c}(p,s)\rangle }{p^{2}-m_{\Xi _{c}}^{2}}  \notag \\
&&\times \langle K(q)\Xi _{c}(p,s)|\Omega _{c}^{\prime }(p^{\prime
},s^{\prime })\rangle \frac{\langle \Omega _{c}^{\prime }(p^{\prime
},s^{\prime })|\overline{\eta }|0\rangle }{p^{\prime }{}^{2}-m^{\prime 2}}%
+\ldots ,  \label{eq:SRDecay}
\end{eqnarray}%
where $p^{\prime }=p+q,\ p$ and $q$ are the momenta of the $\Omega _{c},\
\Xi _{c}$ baryons and $K$ meson, respectively. In the last expression $%
m_{\Xi _{c}}$ is the mass of the $\Xi _{c}^{+}$ baryon. The dots in Eq.\ (%
\ref{eq:SRDecay}) stand for contributions of the higher resonances and
continuum states. Note that in principle the ground state $\Omega_c^{0}$
baryon can also be included into the correlation function. However its mass
remain considerably below the threshold $\Xi_c^{+}K^{-}$ and its decay to
the final state $\Xi_c^{+}K^{-}$ is not kinematically allowed.

We introduce the matrix element of the $\Xi_{c}^{+}$ baryon
\begin{equation*}
\langle 0|\eta _{\Xi }|\Xi _{c}(p,s)\rangle =\lambda _{\Xi _{c}}u(p,s),
\end{equation*}%
and define the strong couplings:
\begin{eqnarray}
&&\langle K(q)\Xi _{c}(p,s)|\Omega _{c}^{\prime }(p^{\prime },s^{\prime
})\rangle =g_{\Omega ^{\prime }\Xi K}\overline{u}(p,s)\gamma _{5}u(p^{\prime
},s^{\prime }),  \notag \\
&&\langle K(q)\Xi _{c}(p,s)|\Omega _{c}^{-}(p^{\prime },s^{\prime })\rangle
=g_{\Omega ^{-}\Xi K}\overline{u}(p,s)u(p^{\prime },s^{\prime }).  \notag \\
&&{}
\end{eqnarray}%
Then using the matrix elements of the $\Omega _{c}^{-}$ and $\Omega
_{c}^{\prime }$ baryons, and performing the summation over $s$ and $%
s^{\prime }$ we recast the function $\Pi ^{\mathrm{Phys}}(p,q)$ into the
form:%
\begin{eqnarray}
&&\Pi ^{\mathrm{Phys}}(p,q)=-\frac{g_{\Omega ^{-}\Xi K}\lambda _{\Xi _{c}}%
\widetilde{\lambda }}{(p^{2}-m_{\Xi _{c}}^{2})(p^{\prime }{}^{2}-\widetilde{m%
}^{2})}(\slashed p+m_{\Xi _{c}})  \notag \\
&&\times \left( \slashed p+\slashed q+\widetilde{m}\right) \gamma _{5}+\frac{%
g_{\Omega ^{\prime }\Xi K}\lambda _{\Xi _{c}}\lambda ^{\prime }}{%
(p^{2}-m_{\Xi _{c}}^{2})(p^{\prime }{}^{2}-m^{\prime 2})}  \notag \\
&&\times (\slashed p+m_{\Xi _{c}})\gamma _{5}\left( \slashed p+\slashed %
q+m^{\prime }\right) +\ldots .
\end{eqnarray}%
The double Borel transformation on the variables $p^{2}$ and $p^{\prime 2}$
applied to $\Pi ^{\mathrm{Phys}}(p,q)$ yields
\begin{eqnarray}
&&\mathcal{B}\Pi ^{\mathrm{Phys}}(p,q)=g_{\Omega ^{-}\Xi K}\lambda _{\Xi
_{c}}\widetilde{\lambda }e^{-\widetilde{m}^{2}/M_{1}^{2}}e^{-m_{\Xi
_{c}}^{2}/M_{2}^{2}}  \notag \\
&&\times \left\{ \slashed q\slashed p\gamma _{5}-m_{\Xi _{c}}\slashed %
q\gamma _{5}-\left( \widetilde{m}+m_{\Xi _{c}}\right) \slashed p\gamma
_{5}\right.  \notag \\
&&\left. +\left[ m_{K}^{2}-\widetilde{m}(\widetilde{m}+m_{\Xi _{c}})\right]
\gamma _{5}\right\} +g_{\Omega ^{\prime }\Xi K}\lambda _{\Xi _{c}}\lambda
^{\prime }  \notag \\
&&\times e^{-m^{\prime 2}/M_{1}^{2}}e^{-m_{\Xi _{c}}^{2}/M_{2}^{2}}\left\{ %
\slashed q\slashed p\gamma _{5}-m_{\Xi _{c}}\slashed q\gamma _{5}+\left(
m^{\prime }-m_{\Xi _{c}}\right) \slashed p\gamma _{5}\right.  \notag \\
&&\left. +\left[ m_{K}^{2}-m^{\prime }(m^{\prime }-m_{\Xi _{c}})\right]
\gamma _{5}\right\} ,  \label{eq:CFunc1/2}
\end{eqnarray}%
where $m_{K}^{2}=q^{2}$ is the mass of the $K$ meson, and $M_{1}^{2}$ and $%
M_{2}^{2}$ are the Borel parameters.

As is seen, there are different structures in Eq.\ (\ref{eq:CFunc1/2}),
which can be used to derive the sum rules for the strong couplings. We work
with the structures $\slashed q\slashed p\gamma _{5}$ and $\slashed p\gamma
_{5}$. Separating the relevant terms in the Borel transformation of the
correlation function $\Pi ^{\mathrm{OPE}}(p,\,q)$ computed employing the
quark-gluon degrees of freedom we get:%
\begin{equation}
g_{\Omega ^{-}\Xi K}=\frac{e^{\widetilde{m}^{2}/M_{1}^{2}}e^{m_{\Xi
_{c}}^{2}/M_{2}^{2}}}{\lambda _{\Xi _{c}}\widetilde{\lambda }(m^{\prime }+%
\widetilde{m})}\left[ (m^{\prime }-m_{\Xi _{c}})\mathcal{B}\Pi _{1}^{\mathrm{%
^{\prime}OPE}}-\mathcal{B}\Pi _{2}^{\mathrm{^{\prime}OPE}}\right] ,
\end{equation}%
and
\begin{equation}
g_{\Omega ^{\prime }\Xi K}=\frac{e^{m^{\prime 2}/M_{1}^{2}}e^{m_{\Xi
_{c}}^{2}/M_{2}^{2}}}{\lambda _{\Xi _{c}}\lambda ^{\prime }(m^{\prime }+%
\widetilde{m})}\left[ (\widetilde{m}+m_{\Xi _{c}})\mathcal{B}\Pi _{1}^{%
\mathrm{^{\prime}OPE}}+\mathcal{B}\Pi _{2}^{\mathrm{^{\prime}OPE}}\right] ,
\end{equation}%
where $\Pi _{1}^{\mathrm{^{\prime}OPE}}(p^{2},\,p^{\prime 2})$ and $\Pi
_{2}^{\mathrm{^{\prime}OPE}}(p^{2},\,p^{\prime 2})$ are the invariant
amplitudes corresponding to structures $\slashed q\slashed p\gamma _{5}$ and
$\slashed p\gamma _{5}$, respectively.

The general expressions obtained above contain two Borel parameters $M_1^{2}$
and $M_1^{2}$. But in our analysis we choose
\begin{equation}
M_{1(2)}^{2}= 2M^{2},\,\, M^2=\frac{M_1^{2}M_2^{2}}{M_1^{2}+M_2^{2}},
\end{equation}
which is traditionally justified by a fact that masses of the involved heavy
baryons $\Omega _{c}^{0}$ and $\Xi _{c}^{+}$ are close to each other.

Using the couplings $g_{\Omega ^{-}\Xi K}$ and $g_{\Omega ^{\prime }\Xi K}$
we can easily calculate the width of $\Omega _{c}^{-}\rightarrow \Xi
_{c}^{+}K^{-}$ and $\Omega _{c}^{\prime }\rightarrow \Xi _{c}^{+}K^{-}$
decays. After some computations we obtain:%
\begin{eqnarray}
\Gamma \left( \Omega _{c}^{-}\rightarrow \Xi _{c}^{+}K^{-}\right) &=&\frac{%
g_{\Omega ^{-}\Xi K}^{2}}{8\pi \widetilde{m}^{2}}\left[ (\widetilde{m}%
+m_{\Xi _{c}})^{2}-m_{K}^{2}\right]  \notag \\
&&\times f(\widetilde{m},m_{\Xi _{c}},m_{K}).
\end{eqnarray}%
and%
\begin{eqnarray}
\Gamma \left( \Omega _{c}^{\prime }\rightarrow \Xi _{c}^{+}K^{-}\right) &=&%
\frac{g_{\Omega ^{\prime }\Xi K}^{2}}{8\pi m^{\prime 2}}\left[ (m^{\prime
}-m_{\Xi _{c}})^{2}-m_{K}^{2}\right]  \notag \\
&&\times f(m^{\prime },m_{\Xi _{c}},m_{K}),
\end{eqnarray}
In expressions above the function $f(x,y,z)$ is given as:
\begin{equation*}
f(x,y,z)=\frac{1}{2x}\sqrt{%
x^{4}+y^{4}+z^{4}-2x^{2}y^{2}-2x^{2}z^{2}-2y^{2}z^{2}}.
\end{equation*}

The QCD side of the correlation function $\Pi ^{\mathrm{OPE}}(p,q)$ can be
found by contracting quark fields, and inserting into the obtained
expression relevant propagators. The remaining non-local quark fields $%
\overline{s}_{\alpha }^{a}u_{\beta }^{b}$ have to be expanded using%
\begin{equation*}
\overline{s}_{\alpha }^{a}u_{\beta }^{b}=\frac{1}{4}\Gamma _{\beta \alpha
}^{i}(\overline{s}^{a}\Gamma ^{i}u^{b}),
\end{equation*}%
where $\ \Gamma ^{i}=1,\ \gamma _{5},\ \gamma _{\mu },\ i\gamma _{5}\gamma
_{\mu },\ \sigma _{\mu \nu }/\sqrt{2}$ is the full set of Dirac matrices.
Sandwiched between the K-meson and vacuum states these terms, as well as
ones generated by insertion of the gluon field strength tensor $G_{\lambda
\rho }(uv)$ \ from quark propagators, give rise to the K-meson's
distribution amplitudes of various quark-gluon contents and twists. Both in
analytical and numerical calculations we take into account the K-meson DAs
up to twist-4 and employ their explicit expressions from Ref.\ \cite%
{Ball:2006wn}.

Apart from parameters in the distribution amplitudes, the sum rules for the
couplings depend also on numerical values of the $\Xi_c^{+}$ baryon's mass
and pole residue. In numerical calculations we utilize
\begin{equation}
m_{\Xi_c}=2467.8^{+0.4}_{-0.6}\ \mathrm{MeV},\ \lambda_{\Xi_c}=0.054 \pm
0.020\ \mathrm{GeV}^3,
\end{equation}
from Refs.\ \cite{Olive:2016xmw} and \cite{Azizi:2016dmr}, respectively. The
Borel and threshold parameters for the decay of a baryon are chosen exactly
as in computations of its mass. The auxiliary parameters $\beta$ in the
interpolating currents of $\Omega_c^{0}$ and $\Xi_c^{+}$ baryons are taken
equal to each other and varied within the limits $\cos\theta \in [-0.75,
-0.3]$ and $[0.3, 0.75]$, which are a little bit extended compared to the
mass rules (see, Eq.\ (\ref{eq:Beta})).

Numerical calculations lead to the following values for the strong couplings%
\begin{equation}
g_{\Omega ^{-}\Xi K}=0.48\pm 0.09,\ \ g_{\Omega ^{\prime }\Xi K}=6.18\pm
1.92.
\end{equation}

The predictions for the width of $\Omega _{c}^{-}\to \Xi _{c}^{+}K^{-}$ and $%
\Omega _{c}^{\prime }\to \Xi_{c}^{+}K^{-} $ decays are collected in Table\ %
\ref{tab:Lit2} and compared with the LHCb data and results of other
theoretical works.

\begin{widetext}

\begin{table}[tbp]
\begin{tabular}{|c|c|c|c|c|c|}
\hline\hline
$\Omega_c^{0}$ & $\Omega_c(3000)^{0}$ & $\Omega_c(3050)^{0}$ &  $\Omega_c(3066)^{0}$ &
$\Omega_c(3090)^{0}$ & $\Omega_c(3119)^{0}$  \\
 & ($\mathrm {MeV}$) & ($\mathrm {MeV}$) & ($\mathrm MeV$) & ($\mathrm {MeV}$) & ($\mathrm {MeV}$) \\ \hline\hline
Ref.\ \cite{LHCb} & $4.5 \pm 0.6 \pm 0.3$ & $0.8 \pm 0.2 \pm 0.1$ & $3.5 \pm 0.4 \pm 0.2$ & $8.7 \pm 1.0 \pm 0.8$ & $1.1 \pm 0.8 \pm 0.4$  \\ \hline
This work & $4.7 \pm 1.2$ & $0.6 \pm 0.2$ & $6.4 \pm 1.7 $ & $-$ & $1.9 \pm 0.6$  \\ \hline
Ref.\ \cite{Wang:2017hej} & $4.18$ & $1.12$ & $2.0$ & $4.71$ & $0.074$  \\ \hline
Ref.\ \cite{Chen:2017}& $-$ & $2.7$ & $3.3$ & $8.8$ & $0.7$  \\
\hline\hline
\end{tabular}%
\caption{The theoretical predictions and experimental data for  width of the $  \Omega_c^{0}$ states. }
\label{tab:Lit2}
\end{table}

\end{widetext}


\section{ $\Omega _{c}^{\star -}\to \Xi _{c}^{+}K^{-}$, $\Omega
_{c}^{\star\prime }\to \Xi _{c}^{+}K^{-}$ and $\Omega _{c}^{\star \prime
}\to \Xi_{c}^{\prime +}K^{-}$ decays}

\label{sec:Decays3/2}
The decays of the spin-$3/2$ baryons $\Omega _{c}^{\star \prime }$ and $%
\Omega _{c}^{\star -}$ to $\Xi _{c}^{+}\,K^{-}$ can analyzed as it has been
done for the spin-$1/2$ baryons. Additionally, we take into account, that
the radially excited $\Omega _{c}^{\star \prime }$ baryon can decay through
the channel $\Omega _{c}^{\star \prime }\rightarrow \Xi _{c}^{\prime
+}\,K^{-}$, as well. The $\Xi _{c}^{\prime +}$ is spin-$1/2$ ground-state
baryon, and it belongs to sextet part of the heavy baryons. Its
interpolating current should be symmetric under exchange of the two light
quarks. In this section we consider these three decay processes.

Again we start from the same correlation function, but with the current $%
\eta (x)$ replaced by $\eta _{\mu }(x)$:
\begin{equation}
\Pi _{\mu }(p,q)=i\int d^{4}xe^{ipx}\langle K(q)|\mathcal{T}\{\eta _{\Xi
_{c}}(x)\overline{\eta }_{\mu }(0)\}|0\rangle .
\end{equation}%
We define the strong couplings $g_{\Omega ^{\star -}\Xi K}$ and $g_{\Omega
^{\star \prime }\Xi K}$ through matrix elements
\begin{eqnarray}
&&\langle K(q)\Xi _{c}(p,s)|\Omega _{c}^{\ast -}(p^{\prime },s^{\prime
})\rangle =g_{\Omega ^{\star -}\Xi K}\overline{u}(p,s)\gamma _{5}u_{\alpha
}(p^{\prime },s^{\prime })q^{\alpha },  \notag \\
&&\langle K(q)\Xi _{c}(p,s)|\Omega _{c}^{\ast \prime }(p^{\prime },s^{\prime
})\rangle =g_{\Omega ^{\star \prime }\Xi K}\overline{u}(p,s)u_{\alpha
}(p^{\prime },s^{\prime })q^{\alpha }.  \notag \\
&&{}
\end{eqnarray}%
and for $\Pi _{\mu }^{\mathrm{Phys}}(p,q)$ obtain the following expression:
\begin{eqnarray}
&&\Pi _{\mu }^{\mathrm{Phys}}(p,q)=\frac{g_{\Omega ^{\star -}\Xi K}\lambda
_{\Xi _{c}}\widetilde{\lambda }}{(p^{2}-m_{\Xi _{c}}^{2})(p^{\prime }{}^{2}-%
\widetilde{m}^{2})}q^{\alpha }(\slashed p+m_{\Xi _{c}})\gamma _{5}  \notag \\
&&\times \left( \slashed p+\slashed q+\widetilde{m}\right) F_{\alpha \mu }(%
\widetilde{m})\gamma _{5}-\frac{g_{\Omega ^{\star \prime }\Xi K}\lambda
_{\Xi _{c}}\lambda ^{\prime }}{(p^{2}-m_{\Xi _{c}}^{2})(p^{\prime
2}-m^{\prime 2})}  \notag \\
&&\times q^{\alpha }(\slashed p+m_{\Xi _{c}})\left( \slashed p+\slashed %
q+m^{\prime }\right) F_{\alpha \mu }(m^{\prime })+\ldots,
\end{eqnarray}%
where we have used the shorthand notation
\begin{eqnarray}
&&F_{\alpha \mu }(m)=g_{\alpha \mu }-\frac{1}{3}\gamma _{\alpha }\gamma
_{\mu }-\frac{2}{3m^{2}}(p_{\alpha }+q_{\alpha })(p_{\mu }+q_{\mu })  \notag
\\
&&+\frac{1}{3m}\left[ (p_{\alpha }+q_{\alpha })\gamma _{\mu }-(p_{\mu
}+q_{\mu })\gamma _{\alpha }\right] .
\end{eqnarray}%
For the Borel transformation of $\Pi _{\mu }^{\mathrm{Phys}}(p,q)$ we get%
\begin{eqnarray}
&&\mathcal{B}\Pi _{\mu }^{\mathrm{Phys}}(p,q)=g_{\Omega ^{\star -}\Xi
K}\lambda _{\Xi _{c}}\widetilde{\lambda }e^{-\widetilde{m}%
^{2}/M_{1}^{2}}e^{-m_{\Xi _{c}}^{2}/M_{2}^{2}}q^{\alpha }  \notag \\
&&\times (\slashed p+m_{\Xi _{c}})\gamma _{5}\left( \slashed p+\slashed q+%
\widetilde{m}\right) F_{\alpha \mu }(\widetilde{m})\gamma _{5}-g_{\Omega
^{\star \prime }\Xi K}\lambda _{\Xi _{c}}\lambda ^{\prime }  \notag \\
&&\times e^{-m^{\prime 2}/M_{1}^{2}}e^{-m_{\Xi _{c}}^{2}/M_{2}^{2}}q^{\alpha
}(\slashed p+m_{\Xi _{c}})\left( \slashed p+\slashed q+m^{\prime }\right)
F_{\alpha \mu }(m^{\prime }).  \notag \\
&&{}
\end{eqnarray}%
To extract the sum rules we choose the structures $\slashed q\slashed %
p\gamma _{\mu }$ and $\slashed qq_{\mu }$. \ The same structures should be
isolated in $\mathcal{B}\Pi _{\mu }^{\mathrm{QCD}}(p,q)$ and matched with
ones from $\mathcal{B}\Pi _{\mu }^{\mathrm{Phys}}(p,q)$. The final formulas
for the strong couplings are rather lengthy, therefore we refrain from
providing their explicit expressions.

The knowledge of the strong couplings allows us to find the widths of the
corresponding decay channels. Thus, the width of the $\Omega _{c}^{\ast
-}\rightarrow \Xi _{c}^{+}K^{-}$ decay can be obtained as%
\begin{eqnarray}
\Gamma (\Omega _{c}^{\ast -} &\rightarrow &\Xi _{c}^{+}K^{-})=\frac{%
g_{\Omega ^{\star -}\Xi K}^{2}}{24\pi \widetilde{m}^{2}}\left[ (\widetilde{m}%
-m_{\Xi _{c}})^{2}-m_{K}^{2}\right]  \notag \\
&&\times f^3(\widetilde{m},m_{\Xi _{c}},m_{K}).
\end{eqnarray}%
whereas for $\Gamma (\Omega _{c}^{\ast \prime }\rightarrow \Xi _{c}^{+}K^{-})$
we get
\begin{eqnarray}
\Gamma (\Omega _{c}^{\ast \prime } &\rightarrow &\Xi _{c}^{+}K^{-})=\frac{%
g_{\Omega ^{\star \prime }\Xi K}^{2}}{24\pi m^{\prime 2}}\left[ (m^{\prime
}+m_{\Xi _{c}})^{2}-m_{K}^{2}\right]  \notag \\
&&\times f^3(m^{\prime },m_{\Xi _{c}},m_{K}),
\end{eqnarray}

In order to find $g_{\Omega ^{\star \prime }\Xi ^{\prime }K}$ corresponding
to the vertex $\Omega _{c}^{\star \prime }\Xi _{c}^{\prime +}K^{-}$, we
again use the correlation function $\Pi _{\mu }(p,q) $, but with the current
$\eta _{\Xi _{c}^{\prime }}$,
\begin{eqnarray}
&&\eta _{\Xi _{c}^{\prime }}=-\frac{1}{\sqrt{2}}\epsilon ^{abc}\left\{
\left( u^{aT}Cc^{b}\right) \gamma _{5}s^{c}+\beta \left( u^{aT}C\gamma
_{5}c^{b}\right) s^{c}\right.  \notag \\
&&\left. -\left[ \left( c^{aT}Cs^{b}\right) \gamma _{5}u^{c}+\beta \left(
c^{aT}C\gamma _{5}s^{b}\right) u^{c}\right] \right\} .
\end{eqnarray}%
We skip details and provide below only final expression for the double Borel
transformation of the term $\sim \slashed q\slashed p\gamma _{\mu }$ in $%
\widetilde{\Pi }_{\mu }^{\mathrm{Phys}}(p,q)$, which is utilized to derive
the required sum rule%
\begin{eqnarray}
&&\mathcal{B}\widetilde{\Pi }_{\mu }^{\mathrm{Phys}}(p,q)=-\frac{g_{\Omega
^{\star \prime }\Xi ^{\prime }K}\lambda _{\Xi _{c}^{\prime }}\lambda
^{\prime }}{6m^{\prime }}e^{-m^{\prime 2}/M_{1}^{2}}e^{-m_{\Xi _{c}^{\prime
}}^{2}/M_{2}^{2}}  \notag \\
&&\times \left[ \left( m^{\prime }+m_{\Xi _{c}^{\prime }}\right)
^{2}-m_{K}^{2}\right] \slashed q\slashed p\gamma _{\mu }.
\label{eq:BorelXiprime}
\end{eqnarray}%
In Eq.\ (\ref{eq:BorelXiprime}) $m_{\Xi _{c}^{\prime }}$ and $\lambda _{\Xi
_{c}^{\prime }}$ are the $\Xi _{c}^{\prime +}$ baryon's mass and pole
residue, respectively.

The coupling $g_{\Omega ^{\star \prime }\Xi ^{\prime }K}$ and widths of the
decay $\Omega _{c}^{\ast \prime }\rightarrow \Xi _{c}^{\prime +}K^{-}$ are
given be the expressions:%
\begin{equation*}
g_{\Omega ^{\prime }\Xi ^{\prime }K}=-\frac{e^{m^{\prime
2}/M_{1}^{2}}e^{m_{\Xi _{c}^{\prime }}^{2}/M_{2}^{2}}6m^{\prime}}{\lambda
_{\Xi _{c}^{\prime }}\lambda ^{\prime }\left[ (m^{\prime }+m_{\Xi
_{c}^{\prime }})^{2}-m_{K}^{2}\right] }\mathcal{B}\widetilde{\Pi }^{\mathrm{%
OPE}},
\end{equation*}%
and%
\begin{eqnarray}
\Gamma (\Omega _{c}^{\star \prime } &\rightarrow &\Xi _{c}^{\prime +}K^{-})=%
\frac{g_{\Omega ^{\star \prime }\Xi ^{\prime }K}^{2}}{24\pi m^{\prime 2}}%
\left[ (m^{\prime }+m_{\Xi _{c}^{\prime }})^{2}-m_{K}^{2}\right] \nonumber \\
&&\times f^3(m^{\prime },m_{\Xi _{c}^{\prime }},m_{K}).
\end{eqnarray}

In numerical computations for the mass and residue of the $\Xi _{c}^{\prime
+}$ baryon we use
\begin{equation}
m_{\Xi _{c}^{\prime }}=2575.6\pm 3.1\ \mathrm{MeV},\ \lambda _{\Xi
_{c}^{\prime }}=0.055\pm 0.016\ \mathrm{GeV}^{3},
\end{equation}%
which are borrowed from Refs.\ \cite{Olive:2016xmw} and \cite{Wang:2009cr},
respectively.

Numerical computations for the strong couplings yield (in $\mathrm{GeV}^{-1}$):
\begin{eqnarray}
&&g_{\Omega ^{\star -}\Xi K}=12.59\pm 1.78,\ \ g_{\Omega ^{\star \prime }\Xi
K}=0.75\pm 0.20,\   \notag \\
&&g_{\Omega ^{\star \prime }\Xi ^{\prime }K}=1.21\pm 0.41.
\end{eqnarray}%
For the decay widths we get:
\begin{eqnarray}
&&\Gamma (\Omega _{c}^{\star -}\rightarrow \Xi _{c}^{+}K^{-})=0.6\pm 0.2\,\
\mathrm{MeV},  \notag \\
&&\Gamma (\Omega _{c}^{\star \prime }\rightarrow \Xi _{c}^{+}K^{-})=1.3\pm
0.4\,\ \mathrm{MeV},  \notag \\
&&\Gamma (\Omega _{c}^{\star \prime }\rightarrow \Xi _{c}^{\prime
+}K^{-})=0.6\pm 0.2\,\ \mathrm{MeV}.
\end{eqnarray}%
Obtained predictions for widths of the $\Omega _{c}^{\star -}$ and $\Omega
_{c}^{\star \prime }$ baryons are shown in Table \ref{tab:Lit2}: for $\Omega
_{c}^{\star \prime }$ we present there a sum of its two possible decay
channels.


\section{Discussion and Concluding remarks}

\label{sec:Concl}
In the present work we have investigated the newly discovered $\Omega_c^{0}$
baryons by means of QCD sum rule method. We have calculated masses and pole
residues of the ground-state and first orbitally and radially excited $%
\Omega_c^{0}$ baryons with the spin-$1/2$ and -$3/2$. To this end, we have
employed two-point QCD sum rule method and started from the ground-state
baryons. We have derived required sum rules for $m_{\Omega_c}$ and $%
\lambda_{\Omega_c}$ using two different structures in the relevant
correlation functions. The masses and residues of the ground-states have
been treated as input information in the sum rules obtained to evaluate
parameters of the first orbitally excited baryons. The same manipulations
have been repeated also in the case of the radially excited states.

The predictions for the masses and residues obtained in the present work
almost coincide with results of our previous paper \cite{Agaev:2017jyt}
excluding numerically small modifications in parameters of the radially
excited baryons. This may be expected, because in the present work we have
employed more sophisticated iterative scheme. Nevertheless, assignments for $%
\Omega_c^{0}$ made in Ref.\ \cite{Agaev:2017jyt} remain valid here, as well
(see, Table\ \ref{tab:Lit1}).

The widths of the $\Omega_c^{0} \to \Xi_c^{+}K^{-}$ decays, calculated in
the context of the QCD full LCSR method, have allowed us to confirm an
essential part of our previous conclusions. Thus, the mass and width of the $%
(1P,\,1/2^{-})$ and $(2S,\,3/2^{+})$ states are in a nice agreements with
the same parameters of the $\Omega_c(3000)$ and $\Omega_c(3119)$ baryons,
respectively. The mass of the orbitally excited state $(1P,\,3/2^{-})$ is
close to $\Omega_c(3050)$. But it may be considered also as the $%
\Omega_c(3066)$ baryon. A decisive argument in favor of $\Omega_c(3050)$ is
the width of the state $(1P,\,3/2^{-})$, which is in excellent agreement
with LHCb measurements. As a result, we do not hesitate to confirm our
previous assignment of $\Omega_c(3050)$ to be the baryon with $J^{P}=3/2^{-}$%
. Situation with the orbitally excited state $(2S,\,1/2^{+})$ is not quite
clear. In fact, its mass and width, allow one to interpret it either as $%
\Omega_c(3066)$ or $\Omega_c(3090)$. We have kept in Tables\ \ref{tab:Lit1}
and \ref{tab:Lit2} our previous classification of the $(2S,\,1/2^{+})$ state
as the $\Omega_c(3066)$ baryon, but its interpretation as $\Omega_c(3090)$
is also legitimate.

The masses of the excited $\Omega_c^{0}$ baryons were predicted in
theoretical literature long before the recent LHCb data. Most of them were
made in the framework of different quark models (see, for example Refs.\
\cite{Ebert:2011kk,Valcarce:2008dr,Vijande:2012mk,Shah:2016nxi}). Within the
two-point QCD sum rule method problems of the $\Omega_c^{0}$ baryons were
addressed in Refs.\ \cite{Wang:2007sqa,Wang:2008hz,Wang:2009cr,Azizi:2015ksa}%
, where the masses of the ground-state and excited $\Omega_c^{0}$ were
found. Obtained in Refs.\ \cite{Azizi:2015ksa} mass of $\Omega_c^{0}$ baryon
with $J^{P}=3/2^{-}$
\begin{equation}
m_{\Omega_c^{\star}}=3080 \pm 120\ \mathrm{GeV}
\end{equation}
within errors agrees both with LHCb data and our present result for $(1P,\
3/2^{-})$ state.

After discovery of the LHCb Collaboration, parameters of new states in the
context of QCD sum rule approach have been also investigated in Refs.\ \cite%
{Wang:2017hej,Aliev:2017led}. In Ref.\ \cite{Wang:2017hej} all of five
states have been considered as negative-parity baryons, whereas in Ref.\
\cite{Aliev:2017led} only two of them have been classified as
negative-parity states. But lack of information about widths of $%
\Omega_c^{0} $ makes incomplete comparison of their results with available
LHCb data.

A situation around excited $\Omega_c^{0}$ states remains controversial and
unclear. Additional efforts of experimental collaborations are necessary to
explore $\Omega_c^{0}$ states, mainly to fix their spin-parities.

\section*{ACKNOWLEDGEMENTS}

K.~A.~ thanks Do\v{g}u\c{s} University for the partial financial support
through the grant BAP 2015-16-D1-B04. The work of H.~S. was supported partly
by BAP grant 2017/018 of Kocaeli University.

\appendix*


\section{ The correlation functions and quark propagators}

\label{sec:App}
\renewcommand{\theequation}{\Alph{section}.\arabic{equation}} The
correlation function for the spin $1/2$ baryons

\begin{equation*}
\Pi (p)=i\int d^{4}xe^{ipx}\langle 0|\mathcal{T}\{\eta (x)\overline{\eta }%
(0)\}|0\rangle ,
\end{equation*}%
in terms of the quark propagators takes the following form:
\begin{widetext}
\begin{eqnarray}
&&\Pi ^{\mathrm {OPE}}(p)=\int d^{4}xe^{ipx}\frac{\epsilon \epsilon ^{\prime
}}{4}\left\{ -\gamma _{5}S_{s}^{ca^{\prime }}(x)\widetilde{S}_{c}^{ab^{\prime
}}(x)S_{s}^{bc^{\prime }}(x)\gamma _{5}+\gamma _{5}S_{s}^{ca^{\prime }}(x)%
\widetilde{S}_{c}^{bb^{\prime }}(x)S_{s}^{ac^{\prime }}(x)\gamma _{5}+\gamma
_{5}S_{s}^{cb^{\prime }}(x)\widetilde{S}_{c}^{aa^{\prime
}}(x)S_{s}^{bc^{\prime }}(x)\gamma _{5}\right.   \notag \\
&&-\gamma _{5}S_{s}^{cb^{\prime }}(x)\widetilde{S}_{c}^{ba^{\prime
}}(x)S_{s}^{ac^{\prime }}(x)\gamma _{5}+\gamma _{5}S_{s}^{cc^{\prime
}}(x)\gamma _{5}\left[ -\mathrm{Tr}\left[ S_{c}^{aa^{\prime }}(x)\widetilde{S%
}_{s}^{bb^{\prime }}(x)\right] +\mathrm{Tr}\left[ S_{c}^{ab^{\prime }}(x)%
\widetilde{S}_{s}^{ba^{\prime }}(x)\right] -\mathrm{Tr}\left[
S_{s}^{aa^{\prime }}(x)\widetilde{S}_{c}^{bb^{\prime }}(x)\right] \right.
\notag \\
&&\left. +\mathrm{Tr}\left[ S_{s}^{ab^{\prime }}(x)\widetilde{S}%
_{c}^{ba^{\prime }}(x)\right] \right] +\beta \left[ -\gamma
_{5}S_{s}^{ca^{\prime }}(x)\gamma _{5}\widetilde{S}_{c}^{ab^{\prime
}}(x)S_{s}^{bc^{\prime }}(x)+\gamma _{5}S_{s}^{ca^{\prime }}(x)\gamma _{5}%
\widetilde{S}_{c}^{bb^{\prime }}(x)S_{s}^{ac^{\prime }}(x)+\gamma
_{5}S_{s}^{cb^{\prime }}(x)\gamma _{5}\widetilde{S}_{c}^{aa^{\prime
}}(x)S_{s}^{bc^{\prime }}(x)\right.   \notag \\
&&\left. -\gamma _{5}S_{s}^{cb^{\prime }}(x)\gamma _{5}\widetilde{S}%
_{c}^{ba^{\prime }}(x)S_{s}^{ac^{\prime }}(x)-S_{s}^{ca^{\prime }}(x)%
\widetilde{S}_{c}^{ab^{\prime }}(x)\gamma _{5}S_{s}^{bc^{\prime }}(x)\gamma
_{5}+S_{s}^{ca^{\prime }}(x)\widetilde{S}_{c}^{bb^{\prime }}(x)\gamma
_{5}S_{s}^{ac^{\prime }}(x)\gamma _{5}+S_{s}^{cb^{\prime }}(x)\widetilde{S}%
_{c}^{aa^{\prime }}(x)\gamma _{5}S_{s}^{bc^{\prime }}(x)\gamma _{5}\right.
\notag \\
&&\left. -S_{s}^{cb^{\prime }}(x)\widetilde{S}_{c}^{ba^{\prime }}(x)\gamma
_{5}S_{s}^{ac^{\prime }}(x)\gamma _{5}+\gamma _{5}S_{s}^{cc^{\prime }}(x)%
\left[ -\mathrm{Tr}\left[ S_{c}^{aa^{\prime }}(x)\gamma _{5}\widetilde{S}%
_{s}^{bb^{\prime }}(x)\right] +\mathrm{Tr}\left[ S_{c}^{ab^{\prime
}}(x)\gamma _{5}\widetilde{S}_{s}^{ba^{\prime }}(x)\right] -\mathrm{Tr}\left[
S_{s}^{aa^{\prime }}(x)\gamma _{5}\widetilde{S}_{c}^{bb^{\prime }}(x)\right]
\right. \right.   \notag \\
&&\left. +\mathrm{Tr}\left[ S_{s}^{ab^{\prime }}(x)\gamma _{5}\widetilde{S}%
_{c}^{ba^{\prime }}(x)\right] \right] +S_{s}^{cc^{\prime }}(x)\gamma _{5}%
\left[ -\mathrm{Tr}\left[ S_{c}^{aa^{\prime }}(x)\widetilde{S}%
_{s}^{bb^{\prime }}(x)\gamma _{5}\right] +\mathrm{Tr}\left[
S_{c}^{ab^{\prime }}(x)\widetilde{S}_{s}^{ba^{\prime }}(x)\gamma _{5}\right]
-\mathrm{Tr}\left[ S_{s}^{aa^{\prime }}(x)\widetilde{S}_{c}^{bb^{\prime
}}(x)\gamma _{5}\right] \right.   \notag \\
&&\left. \left. +\mathrm{Tr}\left[ S_{s}^{ab^{\prime }}(x)\widetilde{S}%
_{c}^{ba^{\prime }}(x)\gamma _{5}\right] \right] \right] +\beta ^{2}\left[
-S_{s}^{ca^{\prime }}(x)\gamma _{5}\widetilde{S}_{c}^{ab^{\prime }}(x)\gamma
_{5}S_{s}^{bc^{\prime }}(x)+S_{s}^{ca^{\prime }}(x)\gamma _{5}\widetilde{S}%
_{c}^{bb^{\prime }}(x)\gamma _{5}S_{s}^{ac^{\prime }}(x)\right.   \notag \\
&&+S_{s}^{cb^{\prime }}(x)\gamma _{5}\widetilde{S}_{c}^{aa^{\prime
}}(x)\gamma _{5}S_{s}^{bc^{\prime }}(x)-S_{s}^{cb^{\prime }}(x)\gamma _{5}%
\widetilde{S}_{c}^{ba^{\prime }}(x)\gamma _{5}S_{s}^{ac^{\prime
}}(x)+S_{s}^{cc^{\prime }}(x)\left[ \mathrm{Tr}\left[ S_{c}^{ba^{\prime
}}(x)\gamma _{5}\widetilde{S}_{s}^{ab^{\prime }}(x)\gamma _{5}\right]
\right.   \notag \\
&&\left. \left. -\mathrm{Tr}\left[ S_{c}^{bb^{\prime }}(x)\gamma _{5}%
\widetilde{S}_{s}^{aa^{\prime }}(x)\gamma _{5}\right] +\mathrm{Tr}\left[
S_{s}^{ba^{\prime }}(x)\gamma _{5}\widetilde{S}_{c}^{ab^{\prime }}(x)\gamma
_{5}\right] -\mathrm{Tr}\left[ S_{c}^{bb^{\prime }}(x)\gamma _{5}\widetilde{S%
}_{c}^{aa^{\prime }}(x)\gamma _{5}\right] \right] \right\}.
\label{eq:A1}
\end{eqnarray}%
For the correlation function of spin $3/2$ baryons we get:%
\begin{eqnarray}
&&\Pi _{\mu \nu }^{\mathrm {OPE}}(p)=\int d^{4}xe^{ipx}\frac{\epsilon
\epsilon ^{\prime }}{3}\left\{ S_{c}^{ca^{\prime }}(x)\gamma _{\nu }\widetilde{S}%
_{s}^{ab^{\prime }}(x)\gamma _{\mu }S_{s}^{bc^{\prime
}}(x)-S_{c}^{ca^{\prime }}(x)\gamma _{\nu }\widetilde{S}_{s}^{bb^{\prime
}}(x)\gamma _{\mu }S_{s}^{ac^{\prime }}(x)-S_{c}^{cb^{\prime }}(x)\gamma
_{\nu }\widetilde{S}_{s}^{aa^{\prime }}(x)\gamma _{\mu }S_{s}^{bc^{\prime
}}(x)\right.   \notag \\
&&+S_{c}^{cb^{\prime }}(x)\gamma _{\nu }\widetilde{S}_{s}^{ba^{\prime
}}(x)\gamma _{\mu }S_{s}^{ac^{\prime }}(x)+S_{s}^{ca^{\prime }}(x)\gamma
_{\nu }\widetilde{S}_{c}^{ab^{\prime }}(x)\gamma _{\mu }S_{s}^{bc^{\prime
}}(x)-S_{s}^{ca^{\prime }}(x)\gamma _{\nu }\widetilde{S}_{c}^{bb^{\prime
}}(x)\gamma _{\mu }S_{s}^{ac^{\prime }}(x)+S_{s}^{ca^{\prime }}(x)\gamma
_{\nu }\widetilde{S}_{s}^{ab^{\prime }}(x)\gamma _{\mu }S_{c}^{bc^{\prime
}}(x)  \notag \\
&&-S_{s}^{ca^{\prime }}(x)\gamma _{\nu }\widetilde{S}_{s}^{bb^{\prime
}}(x)\gamma _{\mu }S_{c}^{ac^{\prime }}(x)-S_{s}^{cb^{\prime }}(x)\gamma
_{\nu }\widetilde{S}_{c}^{aa^{\prime }}(x)\gamma _{\mu }S_{s}^{bc^{\prime
}}(x)+S_{s}^{cb^{\prime }}(x)\gamma _{\nu }\widetilde{S}_{c}^{ba^{\prime
}}(x)\gamma _{\mu }S_{s}^{ac^{\prime }}(x)-S_{s}^{cb^{\prime }}(x)\gamma
_{\nu }\widetilde{S}_{s}^{aa^{\prime }}(x)\gamma _{\mu }S_{c}^{bc^{\prime
}}(x)  \notag \\
&&+S_{s}^{cb^{\prime }}(x)\gamma _{\nu }\widetilde{S}_{s}^{ba^{\prime
}}(x)\gamma _{\mu }S_{c}^{ac^{\prime }}(x)-S_{s}^{cc^{\prime }}(x)\left[
\mathrm{Tr}\left[ S_{c}^{ba^{\prime }}(x)\gamma _{\nu }\widetilde{S}%
_{s}^{ab^{\prime }}(x)\gamma _{\mu }\right] -\mathrm{Tr}\left[
S_{c}^{bb^{\prime }}(x)\gamma _{\nu }\widetilde{S}_{s}^{aa^{\prime
}}(x)\gamma _{\mu }\right] +\mathrm{Tr}\left[ S_{s}^{ba^{\prime }}(x)\gamma
_{\nu }\widetilde{S}_{c}^{ab^{\prime }}(x)\gamma _{\mu }\right] \right.
\notag \\
&&\left. \left. -\mathrm{Tr}\left[ S_{s}^{bb^{\prime }}(x)\gamma _{\nu }%
\widetilde{S}_{c}^{aa^{\prime }}(x)\gamma _{\mu }\right] \right]
-S_{c}^{cc^{\prime }}(x)\left[ \mathrm{Tr}\left[ S_{s}^{ba^{\prime
}}(x)\gamma _{\nu }\widetilde{S}_{s}^{ab^{\prime }}(x)\gamma _{\mu }\right] -%
\mathrm{Tr}\left[ S_{s}^{bb^{\prime }}(x)\gamma _{\nu }\widetilde{S}%
_{s}^{aa^{\prime }}(x)\gamma _{\mu }\right] \right] \right\}.
\label{eq:A2}
\end{eqnarray}%
In Eqs.\ (\ref{eq:A1}) and (\ref{eq:A2}) $\epsilon \epsilon ^{\prime }=\epsilon ^{abc}\epsilon ^{a^{\prime
}b^{\prime }c^{\prime }}$ and $\widetilde{S}_{s(c)}(x)=CS_{s(c)}^{T}(x)C$.

The quark propagators are important ingredients of sum rules calculations.
Below we provide explicit expressions of the light and heavy quark
propagators in the $x$-representation. For the light $q=u,d,s$ quarks we
have
\begin{eqnarray}
&&S_{q}^{ab}(x)=\frac{i\slashed x}{2\pi ^{2}x^{4}}\delta _{ab}-\frac{m_{q}}{%
4\pi ^{2}x^{2}}\delta _{ab}-\frac{\langle \overline{q}q\rangle }{12}\left(
1-i\frac{m_{q}}{4}\slashed x\right) \delta _{ab}-\frac{x^{2}}{192}%
m_{0}^{2}\langle \overline{q}q\rangle \left( 1-i\frac{m_{q}}{6}\slashed %
x\right) \delta _{ab}  \notag \\
&&-ig_{s}\int_{0}^{1}du\left\{ \frac{\slashed x}{16\pi ^{2}x^{2}}G_{ab}^{\mu
\nu }(ux)\sigma _{\mu \nu }-\frac{iux_{\mu }}{4\pi ^{2}x^{2}}G_{ab}^{\mu \nu
}(ux)\gamma _{\nu }-\frac{im_{q}}{32\pi ^{2}}G_{ab}^{\mu \nu }(ux)\sigma
_{\mu \nu }\left[ \ln \left( \frac{-x^{2}\Lambda ^{2}}{4}\right) +2\gamma
_{E}\right] \right\},
\end{eqnarray}%
where $\gamma _{E}\simeq 0.577$ is the Euler constant and $\Lambda $ is the
QCD scale parameter. We have also used the notation $G_{ab}^{\mu \nu }\equiv
G_{A}^{\mu \nu }t_{ab}^{A},\ A=1,2,\ldots 8$, and $t^{A}=\lambda ^{A}/2$,
with $\lambda ^{A}$ being the Gell-Mann matrices.

The heavy $Q=c,b$ quark propagators we get
\begin{eqnarray}
&&S_{Q}^{ab}(x)=\frac{m_{Q}^{2}}{4\pi ^{2}}\frac{K_{1}\left( m_{Q}\sqrt{%
-x^{2}}\right) }{\sqrt{-x^{2}}}\delta _{ab}+i\frac{m_{Q}^{2}}{4\pi ^{2}}%
\frac{\slashed xK_{2}\left( m_{Q}\sqrt{-x^{2}}\right) }{\left( \sqrt{-x^{2}}%
\right) ^{2}}\delta _{ab}-\frac{g_{s}m_{Q}}{16\pi ^{2}}%
\int_{0}^{1}duG_{ab}^{\mu \nu }(ux)\left[ (\sigma _{\mu \nu }\slashed x+%
\slashed x\sigma _{\mu \nu })\right.  \notag \\
&&\left. \times \frac{K_{1}\left( m_{Q}\sqrt{-x^{2}}\right) }{\sqrt{-x^{2}}}%
+2\sigma _{\mu \nu }K_{0}\left( m_{Q}\sqrt{-x^{2}}\right) \right] .
\end{eqnarray}%
The first two terms above is the free heavy quark propagator in the coordinate
representation, and $K_{n}(z)$ are the modified Bessel functions of the
second kind.
\end{widetext}

\end{document}